\documentclass[fleqn,usenatbib]{mnras}

\usepackage[T1]{fontenc}
\usepackage{ae,aecompl}

\usepackage{graphicx}	
\usepackage{amsmath}	
\usepackage{amssymb}	

\usepackage{verbatim}
\usepackage[autostyle]{csquotes}

\usepackage{newtxtext,newtxmath}

\def\spose#1{\hbox to 0pt{#1\hss}}
\def\lta{\mathrel{\spose{\lower 3pt\hbox{$\mathchar"218$}}
     \raise 2.0pt\hbox{$\mathchar"13C$}}}
\def\gta{\mathrel{\spose{\lower 3pt\hbox{$\mathchar"218$}}
     \raise 2.0pt\hbox{$\mathchar"13E$}}}
\def\magphys{\textsc{magphys}}
\def\cigale{{\sc cigale}}
\def\agnfitter{{\sc agnfitter}}
\def\bagpipes{{\sc bagpipes}}
\def\skirtor{{\sc skirtor}}

\defcitealias{Tasse2021}{Paper I}
\defcitealias{Sabater2021}{Paper II}
\defcitealias{Kondapally2021}{Paper III}
\defcitealias{Duncan2021}{Paper IV}

\title[LoTSS Deep Fields V: Source classifications]{The LOFAR Two-metre Sky Survey:
  Deep Fields Data Release 1. \\
  V. Survey description, source classifications and host
  galaxy properties}

\author[P. N. Best et al.]{
P. N. Best$^{1}$\thanks{E-mail: pnb@roe.ac.uk},
R. Kondapally$^{1}$,
W. L. Williams$^{2,3}$,
R. K. Cochrane$^{4}$,
K. J. Duncan$^{1}$,
C. L. Hale$^{1}$,
P. Haskell$^{5}$,
\newauthor
K. Ma{\l}ek$^{6,7}$,
I. McCheyne$^{8}$,
D. J. B. Smith$^{5}$,
L. Wang$^{9,10}$,
A. Botteon$^{11}$,
M. Bonato$^{11,12,13}$,
M. Bondi$^{11}$,
\newauthor
G. Calistro Rivera$^{14}$,
F. Gao$^{9,10}$,
G. G\"{u}rkan$^{15,16}$,
M. J. Hardcastle$^{5}$,
M. J. Jarvis$^{17,18}$
B. Mingo$^{19}$,
\newauthor
H. Miraghaei$^{20}$,
L. K. Morabito$^{21,22}$,
D. Nisbet$^{1}$,
I. Prandoni$^{11}$,
H. J. A. R{\"o}ttgering$^{2}$,
J. Sabater$^{23,1}$,
\newauthor
T. Shimwell$^{24,2}$,
C. Tasse$^{25,26}$,
R. van Weeren$^{2}$
\\
$^{1}$Institute for Astronomy, University of Edinburgh, Royal Observatory, Blackford Hill, Edinburgh, EH9 3HJ, UK\\
$^{2}$Leiden Observatory, Leiden University, PO Box 9513, 2300 RA Leiden, the Netherlands\\ 
$^{3}$SKA Observatory, Jodrell Bank, Lower Withington, Macclesfield, Cheshire, SK11 9FT, UK\\
$^{4}$Harvard-Smithsonian Center for Astrophysics, 60 Garden St, Cambridge, MA 02138, USA\\
$^{5}$Centre for Astrophysics Research, School of Physics, Astronomy and Mathematics, University of Hertfordshire, College Lane, Hatfield AL10 9AB, UK\\
$^{6}$National Centre for Nuclear Research, Pasteura 7, 02-093 Warsaw, Poland\\
$^{7}$Aix Marseille University, CNRS, CNES, LAM, Marseille, France\\
$^{8}$Astronomy Centre, Department of Physics \& Astronomy, University of Sussex, Brighton, BN1 9QH, UK\\
$^{9}$SRON Netherlands Institute for Space Research, Landleven 12, 9747 AD, Groningen, the Netherlands\\
$^{10}$Kapteyn Astronomical Institute, University of Groningen, Postbus 800, 9700 AV Groningen, the Netherlands\\
$^{11}$INAF - Istituto di Radioastronomia, Via Piero Gobetti 101, I-40129 Bologna, Italy\\
$^{12}$Italian ALMA Regional Centre, Via Piero Gobetti 101, I-40129 Bologna, Italy\\
$^{13}$INAF-Osservatorio Astronomico di Padova, Vicolo dell'Osservatorio 5, I-35122, Padova, Italy \\
$^{14}$European Southern Observatory, Karl-Schwarzschild-Stra{\ss}e 2, 85748 Garching bei M{\"u}nchen\\
$^{15}$Th\"uringer State Observatory, Sternwarte 5, D-07778 Tautenburg, Germany\\
$^{16}$CSIRO Space and Astronomy, ANTF, PO Box 1130, Bentley, WA 6102, Australia\\
$^{17}$Astrophysics, Department of Physics, University of Oxford, Keble Road, Oxford OX1 3RH, UK\\
$^{18}$Department of Physics and Astronomy, University of the Western Cape, Robert Sobukwe Road, 7535 Bellville, Cape Town, South Africa\\
$^{19}$School of Physical Sciences, The Open University, Walton Hall, Milton Keynes, MK7 6AA, UK\\
$^{20}$Research Institute for Astronomy and Astrophysics of Maragha (RIAAM), University of Maragheh, Maragheh, Iran\\
$^{21}$Centre for Extragalactic Astronomy, Department of Physics, Durham University, Durham DH1 3LE, UK\\
$^{22}$Institute for Computational Cosmology, Department of Physics, University of Durham, South Road, Durham DH1 3LE, UK\\
$^{23}$UK Astronomy Technology Centre, Royal Observatory, Blackford Hill, Edinburgh, EH9 3HJ, UK\\
$^{24}$ASTRON, the Netherlands Institute for Radio Astronomy, Postbus 2, 7990 AA, Dwingeloo, the Netherlands\\
$^{25}$GEPI, Observatoire de Paris, CNRS, Universite Paris Diderot, 5 place Jules Janssen, 92190 Meudon, France\\
$^{26}$Department of Physics \& Electronics, Rhodes University, PO Box 94, Grahamstown, 6140, South Africa
}

\date{Accepted XXX. Received YYY; in original form ZZZ}

\pubyear{2021}

\begin{document}
\label{firstpage}
\pagerange{\pageref{firstpage}--\pageref{lastpage}}
\maketitle

\begin{abstract}

Source classifications, stellar masses and star formation rates are
presented for $\approx$80,000 radio sources from the first data
release of the Low Frequency Array Two-metre Sky Survey (LoTSS) Deep
Fields, which represents the widest deep radio survey ever
undertaken. Using deep multi-wavelength data spanning from the
ultraviolet to the far-infrared, spectral energy distribution (SED)
fitting is carried out for all of the LoTSS-Deep host galaxies using
four different SED codes, two of which include modelling of the
contributions from an active galactic nucleus (AGN). Comparing the
results of the four codes, galaxies that host a radiative AGN are
identified, and an optimised consensus estimate of the stellar mass
and star-formation rate for each galaxy is derived. Those galaxies
with an excess of radio emission over that expected from
star formation are then identified, and the LoTSS-Deep sources are
divided into four classes: star-forming galaxies, radio-quiet AGN, and
radio-loud high-excitation and low-excitation AGN. Ninety-five per
cent of the sources can be reliably classified, of which more than two-thirds
are star-forming galaxies, ranging from normal galaxies in
the nearby Universe to highly-starbursting systems at
$z>4$. Star-forming galaxies become the dominant population below
150-MHz flux densities of $\approx$1\,mJy, accounting for 90 per cent
of sources at $S_{\rm 150 MHz} \sim 100 \mu$Jy.  Radio-quiet AGN
comprise $\approx$10 per cent of the overall population. Results are
compared against the predictions of the SKADS and T-RECS radio sky
simulations, and improvements to the simulations are suggested.
\end{abstract}




\section{Introduction}
\label{Sec:intro}

Understanding the formation and evolution of galaxies requires a
detailed knowledge of the baryonic processes that both drive and
quench the process of star formation within galaxies across cosmic
time. In this regard, the faint radio sky provides one of the most
important windows on the Universe, as it offers a direct view onto
three critical (and overlapping) populations of objects: star-forming
galaxies, `radio-quiet' active galactic nuclei (AGN), and low
luminosity radio galaxies \citep[e.g.][]{Padovani2016}.

Arguably the most important observational test for any model of galaxy
formation is measurements of the evolution of the cosmic
star-formation rate density across cosmic time, and the distribution
of that star formation amongst the galaxy population at each redshift,
as a function of stellar mass, galaxy morphology, environment, and
other properties.  These crucial measurements require large, unbiased
samples of star-forming galaxies over a wide range of redshifts. Much
progress has been made in understanding the star-forming galaxy
population, at least out to cosmic noon at $z \sim 2$, using a variety
of star-formation indicators \citep[e.g.][]{Madau2014}. The primary
uncertainty is the effect of dust: by cosmic noon, around 85 per cent
of the total star-formation rate (SFR) density of the Universe is
dust-enshrouded \citep[e.g.][]{Dunlop2017}, and a sub-millimetre
(sub-mm) or far-infrared (far-IR) view of the Universe paints a very
different picture of galaxy properties to that of a population
selected at optical (rest-frame ultraviolet) wavelengths
\citep[e.g.][]{Cochrane2021}. Current far-IR surveys are limited by
sensitivity to the more extreme systems, where contamination of the
far-IR light by AGN emission is also a concern
\citep[e.g.][]{Symeonidis2021}.

Radio emission provides a tool to observe the activity of galaxies in
a manner that is independent of dust. For sources without AGN, the
low-frequency radio emission arises primarily from recent supernova
explosions of massive (young) stars \citep[see reviews
  by][]{Condon1992,Kennicutt1998}, and thus directly traces the
current star-formation rate \citep[unless sufficiently low radio
  frequencies are reached such that free-free absorption becomes
  important; e.g.][]{Schober2017}.  New generation radio
interferometers offer sufficient sensitivity and field-of-view to
survey large samples of star-forming galaxies out to high
redshifts. Crucially, they can also provide sufficient angular
resolution that deep surveys are not generally affected by the source
confusion that limits the capabilities of surveys with sub-mm and
far-IR telescopes such as the {\it Herschel} Space Observatory, for
which the vast majority of sources in deep surveys are blends
\citep[e.g.][]{Oliver2012,Scudder2016}.

Star formation within massive galaxies is widely believed to be
regulated in some manner by AGN, due to the large outflows of energy
associated with the growth of supermassive black holes.  AGN activity
occurs in two fundamental modes \citep[e.g.\ see reviews
  by][]{Heckman2014,Hardcastle2020}. At high accretion rates,
accretion of material on to a black hole is understood to occur
through a `standard' geometrically-thin, optically-thick accretion
disk \citep{Shakura1973}, in which around 10 per cent of the rest-mass
energy of the accreting material is emitted in the form of radiation
(`radiative' or `quasar-like' AGN). These AGN can drive outflowing
winds through thermal or radiation pressure \citep[e.g.][and
  references therein]{Fabian2012}, which may have a substantial effect
on the evolution of the host galaxy. Radiatively-efficient AGN
sometimes possess powerful twin radio jets (`radio-loud' quasars or
their edge-on counterparts, the `high-excitation radio galaxies';
HERGs), and many recent works also suggest that even those that do not
(`radio-quiet' AGN) frequently (or maybe even always) possess weak
radio jets \citep[][and references
  therein]{Jarvis2019,Gurkan2019,Macfarlane2021,Morabito2022}. These
AGN are detectable in deep radio surveys, either due to the weak radio
jets or due to the star formation that can accompany the AGN activity.

At lower accretion rates, typically below about 1 per cent of the
Eddington accretion rate, the nature of the accretion flow on to a
supermassive black hole is believed to change: the accretion flow is
thought to become geometrically thick and radiatively inefficient
\citep{Narayan1994,Narayan1995}. A characteristic feature of these
advection-dominated or radiatively-inefficient accretion flows is that
most of the energy that they release is in the form of two-sided radio
jets (`jet-mode' AGN; also referred to as `low-excitation radio
galaxies'). These jet-mode AGN dominate the radio sky at intermediate
flux densities (above a few mJy), and the radio waveband is by far the
most efficient means of identifying these sources. Jet-mode AGN have
been very well-studied in the nearby Universe
\citep[e.g.][]{Best2012}, where it is now widely accepted that they
play a critical role in the evolution of massive galaxies and
clusters, providing an energy input that counter-balances the
radiative cooling losses of the surrounding hot gas and thus
preventing that gas from cooling and forming stars \citep[see reviews
  by][and references
  therein]{McNamara2007,Fabian2012,Kormendy2013,Heckman2014,Hardcastle2020}. Deeper
radio surveys, probing the faint radio sky, enable these
low-luminosity AGN to be detected and studied to higher redshifts
\citep{Best2014,Pracy2016,Williams2018,Whittam2022}, and hence their
role in the evolution of massive galaxies to be determined across
cosmic time.

Deep radio surveys can therefore offer a unique insight into many
aspects of the galaxy and AGN population. However, to
  extract the maximum science from deep radio surveys, it is
essential that they are carried out in regions of the sky which are
extremely well-studied at other wavelengths across the electromagnetic
spectrum. The ancillary data are required to identify the radio source
host galaxies, to estimate their redshifts, to classify the nature of
the radio emission (star formation vs radiatively-efficient AGN vs
jet-mode AGN) and to determine the physical properties of the host
galaxies (stellar mass, star-formation rate, environment, etc).

Until recently, the state-of-the-art in wide-area deep radio surveys
was the VLA-COSMOS 3\,GHz survey \citep{Smolcic2017}, which used the
Very Large Array (VLA) to cover 2 deg$^2$ of the Cosmic Evolution
Survey (COSMOS) field, arguably the best-studied degree-scale
extragalactic field in the sky. \citet{Smolcic2017b} investigated the
multi-wavelength counterparts of the $\approx$10,000 radio sources
detected, and provided classifications, which then allowed several
further investigations of the radio-AGN and star-forming populations
\citep[e.g.][]{Smolcic2017c,Novak2017,Delvecchio2017,Delhaize2017}.
Nevertheless, even the VLA-COSMOS 3\,GHz survey does not have
sufficient sky area to cover all cosmic environments, and may
therefore suffer from cosmic variance effects, as well as having
limited source statistics at the highest redshifts. The on-going
MeerKAT International GigaHertz Tiered Extragalactic Exploration
(MIGHTEE) 1.4\,GHz survey aims to extend sky coverage at this depth to
20\,deg$^2$; \citet{Heywood2022} provide an early release, with
\citet{Whittam2022} deriving source classifications for 88 per cent of
the $\approx 5,000$ sources with host galaxy
  identifications over 0.8\,deg$^2$ in the COSMOS field.

The Low Frequency Array \citep[LOFAR;][]{vanHaarlem2013} Two-metre Sky
Survey (LoTSS) Deep Fields have a similar goal at lower frequency. The
first data release (hereafter LoTSS-Deep DR1) was made public in April
2021: the radio data reach rms sensitivity levels $\approx 4$ times
deeper than the wider all-northern-sky LoTSS survey
\citep{Shimwell2017,Shimwell2019,Shimwell2022}, corresponding to
approximately the same effective depth as the VLA-COSMOS 3\,GHz survey
(for a source with typical radio spectral index, $\alpha \approx 0.7$,
where $S_\nu \propto \nu^{-\alpha}$) but over an order of magnitude
larger sky area \citep[][hereafter Papers I and II
  respectively]{Tasse2021,Sabater2021}. An extensive optical and
near-infrared cross-matching process has identified and provided
detailed photometry for over 97 per cent of the $\approx$80,000 radio
sources detected over the central regions of the target fields where
the best ancillary data are available \citep[a combined area of
  25\,deg$^2$;][Paper III]{Kondapally2021}. These data have been used
to provide high-quality photometric redshifts \citep[][Paper
  IV]{Duncan2021}. In this paper, the 5th of the series, these data
are combined with far-IR data to carry out detailed spectral energy
distribution (SED) fits to the multi-wavelength photometry from
ultraviolet (UV) to far-IR wavelengths, using several different SED
fitting codes. Using the results of this analysis, the radio sources
are classified into their different types, and key physical parameters
of the host galaxies, such as their stellar masses and star-formation
rates, are determined.

The layout of the paper is as follows. In Sec.~\ref{sec:lotss_deep}
the LoTSS Deep Fields survey is described: this section outlines the
choice of target fields, and places the first data release in to the
context of the eventual full scope of the
survey. Sec.~\ref{sec:sedfits} then describes the data that will be
used in the paper and outlines the application of the SED fitting
algorithms.  Sec.~\ref{sec:optAGN} describes how the results are used
to identify the (radiative-mode) AGN within the sample.  The results
of the different SED fitting algorithms are compared in
Sec.~\ref{sec:consensus}, and used to define consensus measurements
for the stellar mass and star-formation rate of each host
galaxy. Combining this information with the radio data,
Sec.~\ref{sec:radAGN} then describes the identification of
radio-excess AGN. Sec.~\ref{sec:classes} summarises the final
classifications of the objects in the sample, and investigates the
dependence of these on radio flux density, luminosity, stellar mass
and redshift. In Sec.~\ref{sec:modelcomp} the results are compared
against the predictions of the most widely-used radio sky simulations,
and suggestions made for improvements to those simulations. Finally,
conclusions are drawn in Sec.~\ref{sec:concs}. The classifications
derived are released in electronic form and are used for detailed
science analysis in several further papers \citep[][and
  others]{Smith2021,Bonato2021,Kondapally2022,McCheyne2022,Mingo2022,Cochrane2023}.

Throughout the paper, cosmological parameters are taken to be
$\Omega_m = 0.3$, $\Omega_{\Lambda} = 0.7$ and $H_0 = 70$ km s$^{-1}$
Mpc$^{-1}$, and the \citet{Chabrier2003} initial mass function is adopted.

\section{The LoTSS Deep Fields}
\label{sec:lotss_deep}

\subsection{LOFAR observations of the LoTSS Deep Fields}
\label{sec:lotss_deep_rad}

The International LOFAR Telescope \citep{vanHaarlem2013} is a
remarkably powerful instrument for carrying out deep and wide radio
surveys of the extragalactic sky, owing to its high sensitivity, high
angular resolution (6 arcsec at 150\,MHz when using only Dutch
baselines, improving to 0.3 arcsec with the international stations
included), and in particular its wide field-of-view. The primary beam
full-width at half-maximum (FWHM) of the Dutch LOFAR stations is 3.8
degrees at 150\,MHz, giving a field-of-view of more than 10\,deg$^2$
in a single pointing. International stations have a larger collecting
area and a correspondingly smaller beam: 2.5\,deg FWHM; 4.8\,deg$^2$
field-of-view. The LoTSS survey
\citep{Shimwell2017,Shimwell2019,Shimwell2022} is exploiting LOFAR's
capabilities by observing the entire northern sky, with a target rms
depth of below 100$\mu$Jy\,beam$^{-1}$ at favourable declinations (the
non-steerable nature of the LOFAR antennas means that sensitivity
decreases at lower elevations). Nevertheless, LoTSS only scratches the
surface of the depth that radio surveys with LOFAR are capable of
reaching. LoTSS provides an excellent census of the radio-loud AGN
population which dominates the bright and intermediate radio sky, but
samples only the brighter end of the radio-quiet AGN and star-forming
galaxy populations which become dominant as the LoTSS flux density
limit is approached.

The LoTSS Deep Fields provide a complementary deeper survey, aiming to
reach a noise level of 10-15\,$\mu$Jy\,beam$^{-1}$ over a sky area of
at least 30\,deg$^2$. LoTSS-Deep is designed to have the sensitivity
to detect Milky-Way-like galaxies out to $z>1$, and galaxies with
star-formation rates of 100$M_{\odot}$\,yr$^{-1}$ to beyond $z=5$
\citep[e.g.][]{Smith2016}, as well as being able to detect typical
radio-quiet quasars right out to redshift 6
\citep{Gloudemans2021}. The sky area makes it possible to: (i) sample
the full range of environments at high redshifts -- for example, it is
expected to include 10 rich proto-clusters at $z > 2$; (ii) include
statistically meaningful samples of rarer objects (such as $z > 5$
starbursts); (iii) build large enough samples of AGN and star-forming
galaxies (over 100,000 of each expected to be detected) to allow
simultaneous division by multiple key properties, such as luminosity,
redshift, stellar mass and environment.

LoTSS-Deep is being achieved through repeated 8-hr LOFAR observations
of the regions of the northern sky with the highest quality
degree-scale multi-wavelength data. The four target fields are the
European Large Area ISO Survey Northern Field 1
\citep[ELAIS-N1;][]{Oliver2000}, the Bo\"otes field
\citep{Jannuzi1999}, the Lockman Hole \citep{Lockman1986} and the
North Ecliptic Pole (NEP); these are described in more detail in
Section~\ref{sec:lotss_deep_opt}.

\begin{table*}
  \begin{center}
    \caption{\label{tab:lotssdeep} Status of observations and imaging
      in LOFAR Deep Fields, including the data released in the LoTSS
      Deep Fields 1st data release (LoTSS-Deep DR1). The area of best
      ancillary data is defined in \citetalias{Kondapally2021}. Quoted
      rms noise levels are those at the centre of the field. The
      marginally lower sensitivity in Bo\"otes compared to the other
      fields is due to its lower declination, and hence lower average
      elevation during the observations. The `number of sources in DR1 full area' quoted
      is over the full catalogues presented in \citetalias{Tasse2021}
      and \citetalias{Sabater2021}, out to the 30 per cent power point
      of the primary beam (ie. over $\sim 25$\,deg$^2$ in each
      field).}
  \begin{tabular}{ccccccccc}
    \hline
    Field & Coordinates & Area of best   & Obs. time & central rms & N$^{\rm o}$ sources & N$^{\rm o}$ sources & Final awarded  & Target  \\ 
    & (J2000) & ancillary data & in DR1 & noise in DR1 & full DR1 & best ancillary & integration    & rms depth \\
    &             & [deg$^2$] & [hrs] & [$\mu$Jy/beam] & area & data area & time [hrs] & [$\mu$Jy/beam] \\ 
   \hline
    ELAIS-N1     & 16 11 00 ~~+54 57 00 & 6.74  & 164 & 19 & 84,862 & 31,610 & 500  & 11 \\
    Bo\"{o}tes   & 14 32 00 ~~+34 30 00 & 8.63  & 80  & 32 & 36,767 & 31,162 & 312  & 16 \\
    Lockman Hole & 10 47 00 ~~+58 05 00 &10.28  & 112 & 22 & 50,112 & 19,179 & 352  & 13 \\
    NEP          & 17 58 00 ~~+66 00 00 & 10.0  & --  & -- & --     & --     & 400 & 13 \\
    \hline
  \end{tabular}
  \end{center}
\end{table*}

\begin{figure*}
    \centering
    \includegraphics[width=0.7\textwidth]{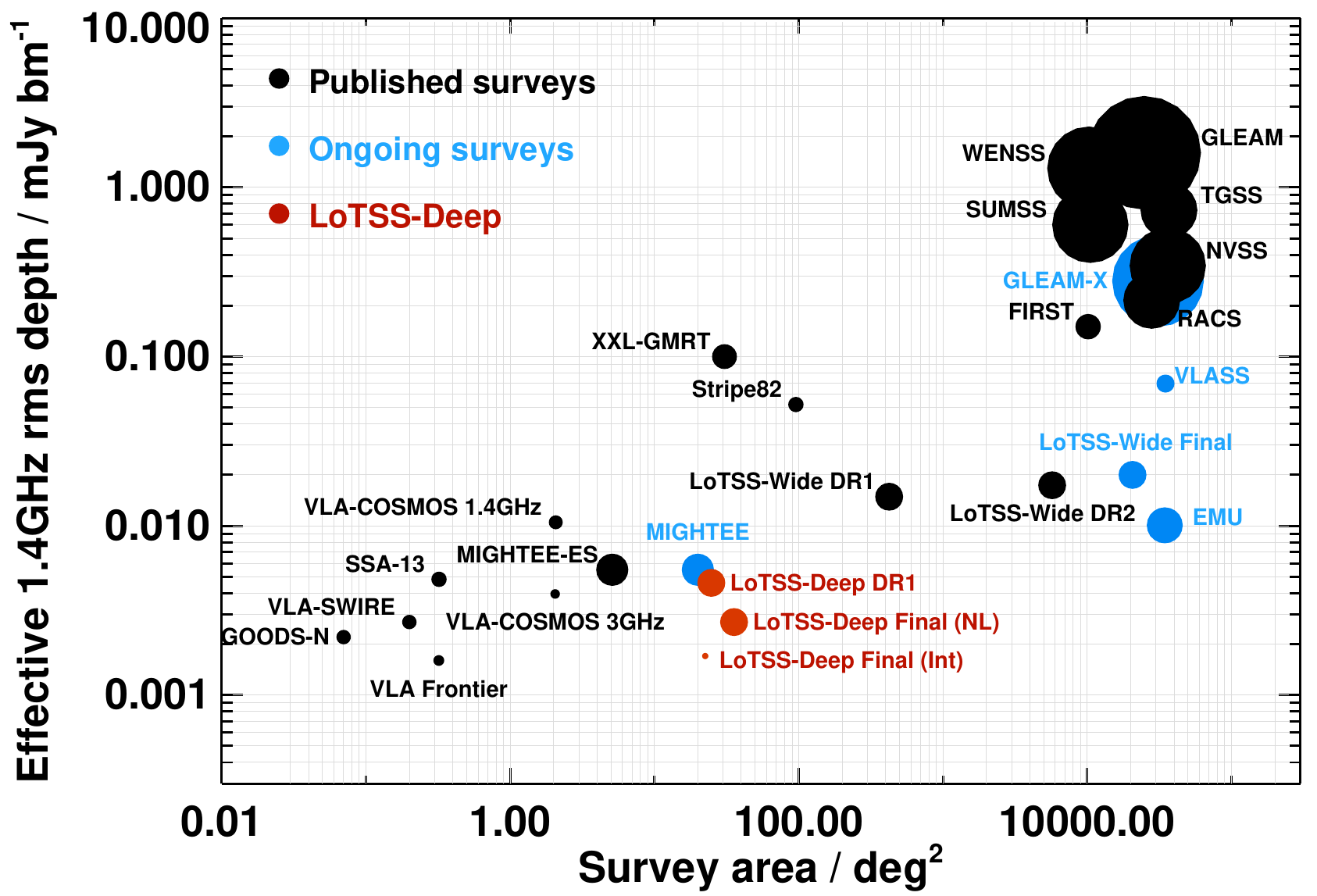}
    \caption{The survey depth, area and angular resolution of the
      LoTSS Deep Fields compared to other existing and on-going radio
      surveys. All survey depths are converted to a 1.4\,GHz
      equivalent rms depth using a spectral index of $\alpha = 0.7$.
      The black points show published surveys, and the blue points
      show on-going surveys. The LoTSS Deep Fields are highlighted in
      red. The size of each symbol indicates the angular resolution of
      the survey, with the symbol area proportional to the beam
      FWHM. For the LoTSS Deep Fields final release, the larger symbol
      indicates the result of including just the Dutch baselines,
      while the smaller symbol shows what should be achievable after
      including the international stations (improved angular
      resolution, additional depth due to the extra collecting area,
      but smaller areal coverage due to the smaller primary beam of
      the international stations). Descriptions of the surveys
      included on the plot (listed from high to low effective rms
      depth) can be found in the following references: GLEAM
      \citep{Wayth2015}; WENSS \citep{Rengelink1997}; TGSS
      \citep{Intema2017}; SUMSS \citep{Mauch2003}; NVSS
      \citep{Condon1998}; GLEAM-X \citep{Hurley2022}; RACS
      \citep{Hale2021}; FIRST \citep{Becker1995}; XXL-GMRT
      \citep{Smolcic2018}; VLASS \citep{Lacy2020}; Stripe82
      \citep{Hodge2011}; LoTSS-Wide \citep{Shimwell2019}; VLA-COSMOS
      1.4\,GHz \citep{Schinnerer2007}; EMU \citep{Norris2011}; MIGHTEE
      \citep[including Early Science -- ES;][]{Heywood2022} SSA-13 \citep{Fomalont2006};
      VLA-COSMOS 3\,GHz \citep{Smolcic2017}; VLA-SWIRE
      \citep{Owen2008}; GOODS-N \citep{Owen2018}; VLA Frontier
      \citep{Heywood2021}.}
    \label{fig:lotssdeep}
\end{figure*}

Table~\ref{tab:lotssdeep} outlines the anticipated final depths of
each field based on awarded observing time. Scaling by depth and area
from radio source counts in shallower LoTSS-Deep observations, the
final LoTSS Deep Fields are expected to detect more than 250,000 radio
sources within the central 35\,deg$^2$, overlapping the best
multi-wavelength data. Figure~\ref{fig:lotssdeep} compares the
sensitivity, field-of-view, and angular resolution of the LoTSS Deep
Fields to other completed and on-going radio surveys. The final LoTSS
Deep Fields dataset will be unrivalled in its combination of depth and
area. The inclusion of the international stations will also provide an
angular resolution which is unmatched by any competitor survey:
indeed, at low frequencies, the LoTSS Deep Fields with international
baselines will remain unique even in the era of the Square Kilometre
Array (SKA).

In order to account for the smaller primary beam of the international
stations, from LOFAR Observing Cycle 14 onwards the pointing positions
for the LoTSS-Deep observations of the Lockman Hole, Bo\"otes and NEP
fields have been dithered around a small mosaic. The mosaics have been
designed to ensure good coverage of the sky area with the best-quality
multi-wavelength data, within the primary beam of the international
stations, while keeping offsets small enough so that there is
negligible loss of sensitivity over this region when imaging with only
Dutch stations.

\subsection{LoTSS-Deep DR1}
\label{sec:lotss_deep_dr1}

This paper considers the radio source catalogues from the first LoTSS
Deep Fields data release. LoTSS-Deep DR1 released the reduced LOFAR
images and catalogues constructed from data taken before October 2018
\citepalias{Tasse2021,Sabater2021}, along with the optical/IR
catalogues and host galaxy identifications \citepalias{Kondapally2021}
and photometric redshifts \citepalias{Duncan2021}. These LoTSS-Deep
DR1 LOFAR observations focused on the ELAIS-N1, Bo\"otes and Lockman
Hole fields, due to the earlier availability of the multi-wavelength
data in those fields. The LoTSS-Deep DR1 LOFAR images included only
the data from the Dutch LOFAR stations, not the international
stations, due to the additional complications associated with
calibrating the long baselines and the associated computing
requirements \citep[see e.g.][for a description of recent advances
  towards a pipeline for international
  stations]{Morabito2022,Sweijen2022}. The data allow an angular
resolution of 6 arcsec to be achieved: higher angular resolution
images will be produced in later data releases.

As shown in Table~\ref{tab:lotssdeep}, the images in LoTSS-Deep DR1
already reach an rms noise level below 20$\mu$Jy\,beam$^{-1}$ at
150\,MHz at the centre of the deepest field (ELAIS-N1), away from
bright sources. Sensitivity decreases with primary beam attenuation
towards the outer regions of the field; dynamic range effects are also
present around bright sources but only a few percent of the image
suffers from significantly increased noise levels due to these
calibration issues \citepalias{Tasse2021,Sabater2021}. Over 170,000
sources are catalogued, with peak flux densities above 5
  times the local rms noise, across the full radio area of the three
  fields; as with all radio catalogues, imcompleteness effects come in
  as the flux limit is approached \citep[see][for an analysis of the
    completeness for AGN and SFGs,
    respectively]{Kondapally2022,Cochrane2023}.  More than 80,000
  sources are catalogued in the central regions with the best
multi-wavelength data \citepalias{Kondapally2021}. As can be seen in
Figure~\ref{fig:lotssdeep}, LoTSS-Deep DR1 broadly matches the depth
of the VLA-COSMOS 3GHz survey but over an order of magnitude larger
sky area; similarly it matches the recent MeerKAT MIGHTEE Early
Release \citep{Heywood2022} in rms depth (the latter being limited by
source confusion owing to its lower angular resolution), but again
over larger area.

\subsection{Multi-wavelength data in the LoTSS Deep Fields}
\label{sec:lotss_deep_opt}

ELAIS-N1, Bo\"otes, Lockman Hole and NEP are the premier large-area
northern extragalactic fields, with vast amounts of telescope time
across the electromagnetic spectrum invested in observing these fields
over the last two decades. Imaging at optical and near-IR wavelengths
reaches 3-4 magnitudes deeper than typical all-sky surveys, allowing
host galaxy identifications for over 97 per cent of the hosts of the
radio sources in LoTSS-Deep DR1 \citepalias{Kondapally2021}
compared to just 73 per cent using all-sky surveys in the LoTSS DR1
release \citep{Williams2019}. Other datasets, such as deep {\it
  Herschel} and {\it Spitzer} data in these fields, are irreplaceable,
and add greatly to the scientific potential: {\it Herschel} data are a
key tool to constrain obscured star-formation rates, while the mid-IR
wavelengths covered by {\it Spitzer} contain the diagnostic emission
from the AGN torus. This range of complementary data makes these
excellent fields to study not only the high-redshift AGN and luminous
star-forming galaxies detected by LOFAR, but also to understand how
this activity sits within the wider cosmological context of the
underlying galaxy population.

As well as their combined benefit of sky area and sample size, each of
the four LoTSS Deep Fields possesses unique characteristics or
datasets which further enhance its specific scientific potential,
whilst complementing each other. The specific data available in each
field are summarised here; a more complete description of the
available data in the ELAIS-N1, Lockman Hole and Bo{\"o}tes fields
(but not NEP, as it was not included in the LoTSS-Deep DR1) can be
found in \citetalias{Kondapally2021}, which also provides the coverage
maps of each survey and the resulting catalogues.

\subsubsection{ELAIS-N1}

ELAIS-N1 has an ideal declination (+55 deg) for LOFAR observations,
and is also a target field for LOFAR's Epoch of Reionisation studies
\citep{Jelic2014}, providing a combined motivation for the
observations. ELAIS-N1 benefits from some of the deepest wide-field
optical, near-IR and mid-IR imaging. It is one of the Medium Deep
Fields from the Panoramic Survey Telescope and Rapid Response Sysytem
(Pan-STARRS-1) survey \citep{Chambers2016}, covering a 7 deg$^2$
field-of-view in the optical $g$,$r$,$i$,$z$,$y$ bands. It is a
Hyper-Suprime-Cam Subaru Strategic Program
\citep[HSC-SSP;][]{Aihara2018} optical deep field, with deep
observations in $g$,$r$,$i$,$z$,$y$ and the narrow-band NB921 over
7.7\,deg$^2$. $u$-band data over this full field are available from
the {\it Spitzer} Adaptation of the Red-sequence Cluster Survey
\citep[SpARCS;][]{Muzzin2009}, and UV data were taken by the Galaxy
Evolution Explorer ({\it GALEX}) space telescope as part of the Deep
Imaging Survey \citep{Martin2005}.  ELAIS-N1 also possesses deep
near-IR imaging in $J$ and $K$ bands from the United Kingdom Infrared
Deep Sky Survey \citep[UKIDSS;][]{Lawrence2007} Deep Extragalactic
Survey (DXS), covering nearly 9\,deg$^2$.

Mid-infrared data were acquired by {\it Spitzer} through both the {\it
  Spitzer} Wide-area Infra-Red Extragalactic survey
\citep[SWIRE;][]{Lonsdale2003} in IRAC channels 1 to 4
(3.6--8.0$\mu$m) over $\sim 10$\,deg$^2$ and the {\it Spitzer}
Extragalactic Representative Volume Survey
\citep[SERVS; ][]{Mauduit2012}, which is around a magnitude deeper at
3.6 and 4.5$\mu$m in the central 2.4\,deg$^2$. Longer wavelength data
in the field have been taken using both {\it Spitzer} (24$\mu$m data
with the Multi-band Imaging Photometer for Spitzer; MIPS) and the {\it
  Herschel} Space Observatory, the latter as part of the {\it Herschel}
Multi-tiered Extragalactic Survey \citep[HerMES;][]{Oliver2012}, one
of the deepest large-area {\it Herschel} surveys.  HerMES observed
ELAIS-N1 at 100$\mu$m, 160$\mu$m, 250$\mu$m, 350$\mu$m and 500$\mu$m.

\subsubsection{Bo\"otes}

The Bo\"otes field is the target of some of the deepest wide-field
optical imaging, in the $B_W$, $R$ and $I$ filters from the NOAO Deep
Wide Field Survey \citep{Jannuzi1999}, in the $z$-band from the
zBo\"otes survey \citep{Cool2007}, and in the $U$ and $Y$ bands from
the Large Binocular Telescope \citep{Bian2013}, all covering around
10\,deg$^2$. The same sky region has been observed in the near-IR $J$,
$H$ and $K$ bands \citep{Gonzalez2010} and using {\it Spitzer} from 3.6 to
8.0$\mu$m as part of the {\it Spitzer} Deep Wide Field Survey
\citep[SDWFS;][]{Ashby2009}. Catalogues of galaxies in the Bo\"otes
field were generated by \citet{Brown2007,Brown2008}. Bo\"otes has also
been observed by {\it Herschel} as part of HerMES, and by {\it Spitzer}-MIPS,
adding far-infrared measurements to the dataset.

In addition to this, Bo\"otes benefits from excellent wide-field X-ray
coverage, including a deep Msec {\it Chandra} survey over the full
9.3\,deg$^2$ field \citep{Masini2020}. The comparison between deep
radio and deep X-ray observations opens many new scientific avenues,
such as investigating the relationship between jet power and accretion
rate in AGN, and determining the black hole accretion rates of
star-forming galaxies to investigate the co-evolution of galaxies and
black holes. Bo\"otes also possesses a vastly higher number of
spectroscopic redshifts than the other northern deep fields, largely
due to the AGN and Galaxy Evolution Survey
\citep[AGES;][]{Kochanek2012}: these are also very valuable for
training photometric redshifts for the radio source population
\citepalias[e.g.][]{Duncan2021}.
 
\subsubsection{Lockman Hole}

Located (like ELAIS-N1) at an ideal declination for LOFAR (+58 deg),
the Lockman Hole is one of the regions of sky with the lowest Galactic
HI column density \citep{Lockman1986}, making it ideal for
extragalactic studies, especially at IR wavelengths due to its low IR
background.  For this reason, the Lockman Hole has been the target of
some of the widest deep coverage in the optical to mid-IR
bands. Optical data in the Lockman Hole has been taken by SpARCS in
$u$,$g$,$r$,$z$ over 13.3\,deg$^2$, and by the Red Cluster Sequence
Lensing Survey \citep[RCSLenS;][]{Hildebrandt2016} in $g$,$r$,$i$,$z$
over 16\,deg$^2$ (albeit not contiguous). As with ELAIS-N1, UV data
have been obtained by the {\it GALEX} Deep Imaging Survey, deep
near-IR $J$ and $K$ band data are available as part of the UKIDSS-DXS
survey (8\,deg$^2$), mid-IR data are available from both SWIRE
(Channels 1--4 over 11\,deg$^2$) and SERVS (3.6 and 4.5\,$\mu$m;
5.6\,deg$^2$) and far-IR data are available over the whole field from
both {\it Spitzer}-MIPS imaging (24$\mu$m) and the {\it Herschel}
HerMES project (100$\mu$m, 160$\mu$m, 250$\mu$m, 350$\mu$m and
500$\mu$m).

The Lockman Hole is arguably the best-studied of the deep fields at
other radio frequencies \citep[e.g.][]{Mahony2016,Prandoni2018,Morganti2021}.
The multi-frequency radio data allow detailed investigations of radio
spectral shapes, identifying peaked, remnant and re-started sources,
and giving a unique insight into the physics and lifecycles of
radio-loud AGN \citep[e.g.][]{Brienza2017,Jurlin2020}.
 
\subsubsection{North Ecliptic Pole}

The North Ecliptic Pole is an interesting field due to its location in
the continuous viewing zone (CVZ) of many space telescopes, including
the {\it JWST}, the \textit{eROSITA} X-ray
mission and \textit{Euclid}. Until very recently, the
multi-wavelength data quality in the NEP was inferior to the other
three LoTSS Deep Fields, but this is rapidly changing.  The NEP is the
location of the \textit{Euclid} Deep Field North which will provide
deep sub-arcsecond near-IR imaging to depths of $H = 26$ over 10
deg$^2$ (and slightly shallower over a wider 20 deg$^2$ region). Such
deep data will enable mass-complete samples to be defined down to
$\sim10^{10} M_{\odot}$ at $z=3$ and normal star-forming galaxies to
be detected out to $z > 6$.  The combination of matched sub-arcsecond
near-IR and radio continuum imaging (with LOFAR's international
baselines) offers a unique opportunity to study the structural
evolution of galaxies, for example comparing the spatial distribution
of star formation (probed by LOFAR) versus stellar mass (probed by
\emph{Euclid}) within galaxies, to cleanly distinguish between
different growth scenarios (e.g. `inside-out' or `outside-in' growth)
over large samples of massive galaxies with $z <1$.

Given these forthcoming datasets, a number of photometric surveys have
been recently undertaken to provide matching observations at other
wavelengths, including the Hawaii Two-0 survey \citep{McPartland2023}.
Additionally, the {\it Euclid}/{\it WFIRST} {\it Spitzer} Legacy
Survey has obtained mid-infrared imaging over the central 10 deg$^2$
of the field using {\it Spitzer} that is $\sim$0.8mag deeper than the
SERVS data available in ELAIS-N1 and Lockman Hole.

As shown in Table~\ref{tab:lotssdeep}, the NEP is not included in
LoTSS-Deep DR1, and hence not included in the analysis of this paper,
as the radio data were not available at the time of the optical
cross-identification. An image from 72-hrs of data is now available
and will be published by \citet{Bondi2023}. Furthermore, as LOFAR
observes two HBA pointings simultaneously, observations of the NEP
field have included a parallel beam centred on the Abell 2255 cluster,
which has also produced an ultra-deep low-frequency image of that
field \citep{Botteon2022}.

\section{Characterising the LoTSS-Deep host galaxies}
\label{sec:sedfits}

\subsection{Optical to mid-IR data}
\label{sec:opticaldata}

For the three fields presented in LoTSS-Deep DR1 (ELAIS-N1, Bo\"otes,
Lockman Hole), \citetalias{Kondapally2021} presented photometric catalogues from
ultraviolet to far-infrared wavelengths. The reader is referred to
that paper for a full description of the catalogues; here, a brief
overview is provided.

For the ELAIS-N1 and Lockman Hole fields, data from UV through to
mid-IR wavelengths were assembled and mosaicked on to a common pixel
scale. Two combined $\chi^2$ signal-to-noise images were then
constructed, one by combining the optical to near-IR bands, and the
other from the {\it Spitzer} 3.6 and 4.5$\mu$m bands; these were
treated separately due to the mis-match in angular resolution between
the ground-based optical-to-near-IR and the {\it Spitzer}
images. Forced aperture photometry was then performed across all bands
using sources detected in each of these stacked images, and the two
catalogues were merged to produce a single consistent photometric
catalogue in each field. Aperture corrections were applied
band-by-band based on curve-of-growth analysis for typical faint
galaxies in order to provide total flux and total magnitude
measurements. The photometry was corrected for galactic extinction
based on the Milky Way E(B-V) extinction map of \citet{Schlegel1998}
and the Milky Way dust extinction law of
\citet{Fitzpatrick1999}. Uncertainties on the photometry were
determined using the variations between a large number of apertures
randomly placed around the fields.

For the Bo\"otes field, forced aperture photometry catalogues already
existed \citep{Brown2007,Brown2008} using magnitude-limited samples
selected in the I-band and the 4.5$\mu$m {\it Spitzer} band. In this case,
these catalogues were used as the starting point, and were merged and
corrected in a similar manner to ELAIS-N1 and Lockman Hole. In all
three fields, the catalogues were then cleaned of low-significance
detections (sources detected in the combined $\chi^2$ image but below
3$\sigma$ significance in each individual band) and cross-talk
artefacts, and those sources in regions around bright stars where
either the cataloguing or the photometry might be unreliable were
flagged, as indicated by the {\sc flag\_clean} parameter. More details
on all of these processes can be found in \citetalias{Kondapally2021}.

These photometric catalogues were then used as the basis for
cross-matching with the LOFAR catalogues. \citetalias{Kondapally2021}
outlines the selection of the studied area for which the
highest-quality multi-wavelength data are available; sources within
this region can be identified using the {\sc flag\_overlap}
parameter. The cross-matching process also involved source
association, such that the catalogued LOFAR sources were combined or
deblended into true physical sources, where necessary. Within these
defined areas, 81,951 physically distinct radio sources were
catalogued over 25.65\,deg$^2$ of sky across the three fields; optical
or near-IR host galaxies were identified for over 97 per cent of these
\citepalias{Kondapally2021}, very much higher than the 73 per cent
found for the wider LoTSS DR1 \citep{Williams2019}.

Photometric redshifts for all of the objects in the field have been
presented in \citetalias{Duncan2021}. These were derived from the UV
to mid-IR data by combining machine learning and template fitting
approaches using a hierarchical Bayesian framework. This method is
shown to provide photometric redshifts which are accurate for both
galaxy populations (out to $z \approx 1.5$) and sources dominated by
AGN emission (out to $z \approx 4$), which is important for the LOFAR
sample. As part of the calibration of the photometric redshifts, small
(typically <5 per cent) offsets in the zero-point magnitudes were
found to improve the accuracy of the template-fit photometric
redshifts. These offsets are discussed further in
Section~\ref{sec:sedcats}.

\subsection{Far-infrared data}
\label{sec:firdata}

The addition of far-IR photometry is described by
\citet{McCheyne2022}, and the reader is referred to that paper for
details. In summary, the far-IR fluxes were measured using XID+
\citep{Hurley2017} which is a Bayesian tool to deblend the flux from
the low resolution {\it Herschel} data into different potential host
galaxies selected from optical/near-IR images. Fluxes were initially
measured as part of the {\it Herschel} Extragalactic Legacy Project
\citep[HELP;][]{Shirley2021}. In HELP, an XID+ prior list of potential
emitters at 24$\mu$m was derived by applying a number of cuts to the
optical-IR galaxy catalogue in order to select the sources most likely
to be bright at 24$\mu$m (those detected both at optical wavelengths
and in the {\it Spitzer} 3.6-8.0$\mu$m bands), and this input list was
used to deblend the 24$\mu$m data. Then, a second prior list was
constructed from those sources with significant 24$\mu$m emission
(above 20$\mu$Jy) and this was used to deblend the {\it Herschel}
data. The posterior distributions for the fluxes derived from XID+
allow the uncertainties to be estimated.

For the LoTSS-Deep catalogue, a cross-match was first made between
each LoTSS-Deep host galaxy position (or its LOFAR position if there
was no host galaxy identification at optical-IR wavelengths) and the
HELP catalogue. If a match was found then the HELP far-IR fluxes were
assigned to the LOFAR source. If no match was found, then XID+ was
re-run following the process above, but with the radio host galaxy
position (or radio position in the case of no host galaxy
identification) added to the prior list: this ensures that the
assignment of zero flux is not simply due to the radio source having
been incorrectly excluded from the prior list.

\subsection{Final catalogues for spectral energy distribution fitting}
\label{sec:sedcats}

In order to ensure consistency and reliability across the different
spectral energy distribution (SED) fitting codes used in this paper,
it was important to ensure that the input dataset was as robust as
possible, and that all photometric errors were uniformly treated.

For each field, a catalogue was produced combining the
(aperture-corrected and Galactic extinction corrected) fluxes from UV
to mid-IR wavelengths with the far-IR fluxes determined by XID+. Next,
the small zero-point magnitude corrections determined during the
photometric redshift fitting were applied: these are tabulated in
Appendix B of \citetalias{Duncan2021}. Specifically, the corrections
derived using the extended Atlas library (referred to as `Brown' in
that paper) were applied; this template set was chosen because it
extended out to the longest IRAC wavelength and also incorporated the
full range of SED types expected within the LoTSS Deep Fields
sample.

The photometry catalogue was then filtered to remove photometric
measurements deemed to be seriously unreliable. These unreliable
measurements were identified as those which were either 2.5 magnitudes
lower, or 1 magnitude higher, than the value predicted by
interpolating the two adjacent filter measurements. These limits were
chosen, following \citet{Duncan2019}, to avoid flagging any reasonable
spectral emission or absorption features, or genuine breaks, while
successfully identifying those measurements that are so discrepant
that they could significantly influence the SED fitting. Around 1 per
cent of the photometric measurements were identified in this way;
these were flagged and not used in the subsequent fitting.

Finally, in order to consistently deal with any residual photometric
errors due to zero-points, aperture corrections or extinction
corrections, 10 per cent of the measured flux was added in quadrature
to all flux uncertainties.  The resultant SED input catalogues for
each field are made available in electronic form through the LOFAR
Surveys website (\url{lofar-surveys.org}).

\subsection{Spectral Energy Distribution fitting}
\label{sec:seds}

Many different codes exist for fitting SEDs to an array of photometric
data points for galaxies and AGN.  Each of these has their own
advantages and disadvantages. \citet{Pacifici2022} recently carried out a
detailed comparison of different codes, finding that they provide
broad agreement in stellar masses, but with more discrepancies in the
star formation rates and dust attenuations derived. In this paper,
four different SED-fitting codes are adopted, and a comparison of the
results between these is used both to derive consensus measurements
for stellar masses and star-formation rates, and to assist with the
classification of the radio source host galaxies.

The `Multi-wavelength Analysis of Galaxy Physical Properties'
\citep[\magphys;][]{daCunha2008} and `Bayesian Analysis of Galaxies
for Physical Inference and Parameter EStimation'
\citep[\bagpipes;][]{Carnall2018,Carnall2019} codes each use energy
balance approaches to fit photometric points from the UV through to
far-IR and sub-mm wavebands. Energy balance implies that the amount of
energy absorbed by dust at optical and UV wavelengths is forced to
match that emitted (thermally) by the dust through the sub-mm and
far-IR. The \magphys\ and \bagpipes\ codes are built on the same
fundamental templates for single stellar populations
\citep{Bruzual2003} but differ in their implementation, in particular
with regard to the parameterisation of the star-formation histories
of the galaxies, the assumed dust models, and the approach to model
optimisation. For high signal-to-noise galaxies the two codes
generally give broadly consistent results (see
Sec.~\ref{sec:consensus}), which previous studies have generally shown
to be accurate \citep[e.g.][]{Hayward2015}. However, neither
\magphys\ nor \bagpipes\ includes AGN emission in its model SEDs, nor
do they account for AGN heating effects when determining energy balance, and
therefore both can give poor fits and unreliable host galaxy
parameters for galaxies with significant AGN emission.

`Code Investigating GALaxy Emission'
\citep[\cigale;][]{Burgarella2005,Noll2009,Boquien2019} is another
broad-band SED-fitting code which uses energy conservation between the
attenuated UV/optical emission and the re-emitted IR/sub-mm emission;
\cigale\ differs from \magphys\ and \bagpipes\ in that it incorporates
AGN models which can account for the direct AGN light contributions
and the infrared emission arising from AGN heating of the dust
\citep[more recent developments also allow for predictions of X-ray
  emission, cf.][]{Yang2020}. The inclusion of AGN models can give
\cigale\ a significant advantage over \magphys\ and \bagpipes\ when
fitting the SEDs of galaxies that have a signficant AGN contribution,
allowing both more robust estimation of host galaxy parameters, and a
mechanism to identify and classify AGN within the sample. However, in
order to allow the additional complications of AGN fitting, for
equivalent (practical) run times \cigale\ is not able to cover the
parameter space of host galaxy properties as finely as \magphys\ and
\bagpipes, leading to potentially less accurate characterisation of
galaxies that do not host AGN.

All of the three codes discussed above adopt the principles of energy
balance. However, if the distribution of ultraviolet light is
spatially disconnected from the dust emission, as is often the case
for very infrared luminous galaxies, then energy balance may not be
valid; indeed, \citet{Buat2019} find for a sample of 17 well-studied
dust-rich galaxies that SED-based UV-optical attenuation estimates
account for less than half of the detected dust emission. This issue
may be particularly pronounced in the presence of AGN, if the AGN
models are not comprehensive enough to properly cover the parameter
space of possible AGN SEDs. To mitigate these issues, the
\agnfitter\ code \citep{Calistro2016} models the SED by independently
fitting four emission components, with each independently normalised
(albeit with a prior that the energy radiated in the infrared must be
at least equal to the starlight energy absorbed by dust at optical/UV
wavelengths): a big blue bump, a stellar population, hot dust emission
from an AGN torus, and colder dust emission. \agnfitter\ can provide
superior fits for objects where energy balance breaks down, and also
for objects with strong AGN components due to its superior modelling
of the big blue bump. However, the lack of energy balance and the
ability of the four components to vary independently can lead to
aphysical solutions, or poorer constraints on the parameters of the
stellar populations (although \citealt{Gao2021} find broadly good
agreement in measured stellar masses and SFRs between codes with and
without energy balance, at least for hyperluminous infrared galaxies).

To maximise the advantages of the different techniques, the LoTSS Deep
Field host galaxies were all modelled using each of \magphys,
\bagpipes, \cigale\ and \agnfitter. Furthermore, for \cigale, two
different sets of AGN models were considered: those of
\citet{Fritz2006} and those of \citet{Stalevski2012,Stalevski2016},
the latter of which were recently incorporated into \cigale\ by
\citet{Yang2020}. The following subsections provide details of the
fitting methodology in each case.

For all SED fitting, the redshift of the source is fixed at the
spectroscopic redshift, $z_{\rm{spec}}$, for the minority of sources
for which this exists (1602, 4039 and 1466 sources in ELAIS-N1,
Bo\"otes and Lockman, respectively). For the other sources, the
redshift is fixed at the median of the first photometric redshift
solution, $z_{\rm 1,median}$. Photometric redshift errors may
introduce errors on the inferred parameters, but for most sources
these are anticipated to be small since the photometric redshifts are
very accurate, with a median scatter of $\Delta z / (1+z) \lta 0.015$
for host-galaxy dominated sources at $z < 1.5$ \citep{Duncan2021}.

\subsubsection{\magphys}
\label{sec:magphys}

The application of \magphys\ to the LoTSS Deep Fields sources is
described by \citet{Smith2021}, and so it is only briefly summarised
here. The stellar population modelling adopts single stellar
population (SSP) templates from \citet{Bruzual2003} and the
two-component (birth cloud plus interstellar medium) dust absorption
model of \citet{Charlot2000}, combined to produce an optical to
near-IR template library of 50,000 SEDs with a range of
exponentially-declining star-formation histories with stochastic
bursts superposed. The dust emission is modelled using a library of
50,000 dust SEDs constructed from dust grains with a realistic range
of sizes and temperatures, including polycyclic aromatic
hydrocarbons. The energy balance criterion is used to combine the two
sets of templates in a physically-viable manner, to produce a model
for the input photometry that stretches from near-UV to sub-mm
wavelengths.

\magphys\ determines the best-fitting SED for every source, returning
the corresponding best-fit physical parameters and their marginalised
probability distribution functions (PDFs). The best-fitting stellar
mass and best-fitting value of the SFR over the last 100\,Myr were
adopted as the stellar mass and SFR respectively; the 100\,Myr
timescale corresponds well to that of the expected radio emission
\citep[e.g.][]{Condon2002}.  For most galaxies, very similar results
are obtained if a shorter period or the current instantaneous SFR are
adopted instead (although results for some individual galaxies can
vary significantly). The 16th and 84th percentile of the PDFs were
adopted as the 1$\sigma$ lower and upper limits respectively. In order
to determine whether the calculated parameters are reliable, the
$\chi^2$ value of the fit was examined: following \citet{Smith2012},
fits for which the determined $\chi^2$ value was above the 99 per cent
confidence limit for the relevant number of photometric bands included
in the fit were flagged as unreliable. As noted by \citet{Smith2021},
many of the objects that fail this test are objects with strong AGN
contributions. On average, 17 per cent of sources across the three
fields were flagged in this way, with ELAIS-N1 giving a significantly
lower fraction (10 per cent), in line with expectations that the
deeper radio data in that field should result in a higher fraction of
star-forming galaxies.

\subsubsection{\bagpipes}
\label{sec:bagpipes}

\bagpipes\ was run on the LoTSS Deep Field sources, making use of the
2016 version of the \citet{Bruzual2003} SSP templates for its stellar
population emission. Nebular emission is computed using the {\sc
  cloudy} photoionization code \citep{Ferland2017}, following
\cite{Byler2017}. {\sc cloudy} is run using each SSP template as the
input spectrum. Dust grains are included using {\sc cloudy}'s `ISM'
prescription, which implements a grain-size distribution and abundance
pattern that reproduces the observed extinction properties for the
Interstellar Medium (ISM) of the Milky Way. A \citet{Calzetti2000}
dust attenuation curve is adopted. Dust emission includes both a hot
dust component from H{\sc II} regions and a grey body component from
the cold, diffuse dust.

A wide dust attenuation prior is adopted, $A_{v}=[0,6]$, which gives
the code the option to fit a high degree of attenuation. The absorbed
energy is re-emitted at infrared wavelengths; the dust SED is
controlled by three key parameters, as described by
\citet{Draine2007}: $U_{\rm{min}}$, the lower limit of the starlight
intensity; $\gamma$, the fraction of stars at $U_{\rm{min}}$; and
$q_{\rm{PAH}}$, the mass fraction of polycyclic aromatic
hydrocarbons. The priors adopted on these parameters are broad, to
allow the model to fit all types of galaxies, including those that are
hot and dusty \citep{Leja2018}: $U_{\rm{min}}=[0,25]$, $\gamma=[0,1]$,
and $q_{\rm{PAH}}=[0,10]$. $\eta$, the multiplicative factor on
$A_{V}$ for stars in birth clouds, is also fitted using the prior
$\eta=[1,5]$. Metallicity is allowed to vary in the range
$Z=[0,2.5]Z_{\odot,\rm{old}}$, where $Z_{\odot,\rm{old}}$ denotes
solar models prior to \cite{Asplund2009}.

The star-formation history (SFH) is parameterised using a double power
law: $\rm{SFR}(t) \propto [(t/\tau)^{\alpha} + (t/\tau)^{-\beta}]^{-1}$
where $\alpha$ is the slope in the region of falling SFR,
and $\beta$ is the slope in the region of rising SFR. $\tau$ relates
to the time at which the SFR peaks.

The code outputs posterior distributions for the fitted parameters
$A_{\rm{V}}$, $U_{\rm{min}}$, $\gamma$, and $q_{\rm{PAH}}$, $\eta$,
the metallicity $Z$, and the SFH parameters $\alpha$, $\beta$ and
$\tau$. Posterior distributions are also derived for the physical
properties of stellar mass, star-formation rate, and specific
star-formation rate, with the median and the 16th and 84th percentiles
being adopted as the best-fit value and the lower and upper 1$\sigma$
errors. The reduced $\chi^2$ of the best-fitting model was also
returned. Objects with a reduced $\chi^2$ above 5 were flagged as
unreliable; this averaged about 9 per cent of sources across the
three fields, again being lowest in ELAIS-N1 and highest in Bo\"otes.

\subsubsection{\cigale}
\label{sec:cigale}

\cigale\ was run on the LoTSS Deep Fields sources in the manner
outlined in \citet{Wang2021} and \citet{Malek2023}. The choices for
the input components for the modelling of the stellar population
largely follow those of \citet{Pearson2018} and
\citet{Malek2018}. Specifically, the star-formation history was
adopted to be a two-component model, with a delayed
exponentially-decaying main star-forming component (SFR$_{\rm delayed}
\propto t e^{-t/\tau}$) plus the addition of a recent starburst. The
\citet{Bruzual2003} SSP templates were adopted for the stellar
emission. The \citet{Charlot2000} dust attenuation model is applied to
the derived SEDs, and energy-balance criteria are used to determine
the quantity of emission to be re-emitted in the infrared. The dust
emission is calculated using the dust emission model of
\citet{Draine2014}, which is an updated version of the
\citet{Draine2007} model and describes the dust as a mixture of
carbonaceous and amorphous silicate grains.

A critical difference between \cigale\ and \magphys/\bagpipes\ is the
inclusion of an AGN component in the \cigale\ models. For the LoTSS
Deep Fields, \cigale\ was run twice, using two different AGN models:
the \citet{Fritz2006} model and the \skirtor\ model of
\citet{Stalevski2012,Stalevski2016}. Both sets of AGN models assume
point-like isotropic emission from a central source, which then
intercepts a toroidal dusty structure close to the AGN. Radiative
transfer models are used to trace the absorption and scattering of the
AGN light by the dust in the torus, and model its re-radiation by the
hot dust. The main differences between the two models are that the
Fritz models adopt a smooth density distribution for the dust grains
and use a 1-D approach, whereas the \skirtor\ models treat the dusty
torus as a two-phase medium with higher density clumps sitting within
a lower density medium and use 3-D radiative transfer. A clumpy dust
distribution was suggested by \citet{Krolik1988} to be necessary to
stop the dust grains being destroyed by the hot surrounding gas.

\cigale\ returns Bayesian estimates of the stellar mass and various
estimates of the recent star-formation rate of the galaxy, along with
estimates of the uncertainties on these parameters. In this work, the
star-formation rate averaged over the last 100\,Myr is adopted, as for
\magphys. \cigale\ also returns a determination of the AGN fraction
for the galaxy (hereafter $f_{\rm AGN,CG-F}$ or $f_{\rm AGN,CG-S}$ for
the Fritz and \skirtor\ models), defined as the fraction of the total
infrared luminosity that is contributed by the AGN dust torus
component. An uncertainty on the AGN fraction is also returned; where
this is larger than the measured fraction, the 1-sigma lower limit on
the AGN fraction is set to zero. Finally, the reduced $\chi^2$ of the
best-fitting model was used to identify unreliable fits, with objects
with a reduced $\chi^2$ above 5 being flagged (3 per cent and 2 per
cent of sources in the Fritz and \skirtor\ models respectively).

\subsubsection{\agnfitter}
\label{sec:agnfitter}

\agnfitter\ provides independent parameterisations for each of the
accretion disk emission (big blue bump), the hot dust torus, the
stellar component and the cooler dust heated by star formation;
details of the parameterisation of these four components are provided
by \citet{Calistro2016}. \agnfitter\ accounts for the effects of
reddening on these emission components but without energy balance
constraints. \agnfitter\ was run on the LoTSS Deep Fields sources
broadly following the implementation of \citet{Williams2018} but using
an expanded set of input models (\agnfitter\ {\sc v2}; Calistro Rivera
et~al., in prep.). The code determines the relative importance of the
four components in a few key wavelength regions, as well as broader
physical parameters including estimates of the star-formation rate and
the stellar mass. In this work, the IR-based estimate of the SFR was
the one adopted.

Following \citet{Williams2018}, an AGN fraction is defined by
considering the contribution of the emission components in the
1-30$\mu$m wavelength range. Note that this is different to the
definition used for \cigale\ which considers the AGN contribution to
the total IR luminosity: as the AGN peaks in the mid-IR, the AGN
fractions derived by \agnfitter\ will typically be larger than those
of \cigale. The AGN fraction was defined as:

\begin{equation}
  f_{\rm AGN,af} = \frac{L_{\rm Torus,1-30}}{L_{\rm Torus,1-30} + L_{\rm SB,1-30} + L_{\rm Gal,1-30}}
\label{eqn:agnfitter}
\end{equation}

\noindent where $L_{\rm Torus,1-30}$, $L_{\rm SB,1-30}$ and $L_{\rm
  Gal,1-30}$ are the luminosities of the hot dust torus, the cooler
dust heated by recent star formation, and the stellar component of the
galaxy, respectively, all between 1 and 30$\mu$m. Note that this
differs slightly from the definition of \citet{Williams2018} through
the inclusion of the stellar component in the denominator; this avoids
a high AGN fraction being determined when the mid-infrared emission is
simply dominated by the light of older stars. The uncertainties on
these luminosities are used to determine the 1$\sigma$ upper and lower
limits to the AGN fraction.

Finally, \agnfitter\ returns a log likelihood for the best-fit model;
the $\approx 3$ per cent of objects whose fits had a log likelihood
below $-$30 were flagged as unreliable \citep[cf.][]{Williams2018}.

\section{Identification of radiative-mode AGN}
\label{sec:optAGN}

A characteristic feature of radiative-mode AGN is a hot accretion
disk, which is being obscured in certain directions by a dusty
structure (the torus). These two structures give rise to a variety of
physical features that can be used to identify the radiative-mode
AGN. The most widely-used of these, where spectroscopic data is
available, is emission line ratios \citep[e.g.][the BPT
  diagram]{Baldwin1981}: the ionising radiation from the hot accretion
disk is significantly harder than that of a young stellar population,
leading to stronger high-excitation forbidden lines. Spectroscopic
information is available for only a small subset of the LoTSS-Deep sources
(5.1, 21.1 and 4.7 per cent in ELAIS-N1, Bo\"otes and Lockman Hole
respectively, with the AGES data in Bo\"otes producing the large
difference between the fields), so this method cannot be used for the
vast majority of the sources. This will change in the coming years due
to the WEAVE-LOFAR survey \citep[][see also
  Sec.~\ref{sec:concs}]{Smith2016} but alternative methods are needed
for AGN identification in the meantime.

The hot dusty torus emits characteristic emission that has been widely
used to identify radiative-mode AGN using mid-IR colours
\citep[e.g.][]{Lacy2004,Stern2005}. Commonly-used selections consider
the four {\it Spitzer} channels centred at 3.6$\mu$m, 4.5$\mu$m,
5.8$\mu$m and 8.0$\mu$m (Channels 1 to 4 respectively); the selection
is based on the premise that the emission from stellar populations
generally declines with increasing wavelength through the mid-IR
(since the mid-IR probes redward of the rest-frame 1.6$\mu$m thermal
peak of the dominant sub-solar stellar population) whereas hot AGN
dust shows a rising spectrum. An equivalent approach uses the WISE
mid-infrared colours \citep[e.g.][]{Wright2010}. The exact
colour-space cuts are generally defined using template tracks for
galaxies and AGN to select regions of colour-space dominated by AGN.

\begin{figure*}
    \begin{tabular}{cc}
    \includegraphics[width=7.5cm]{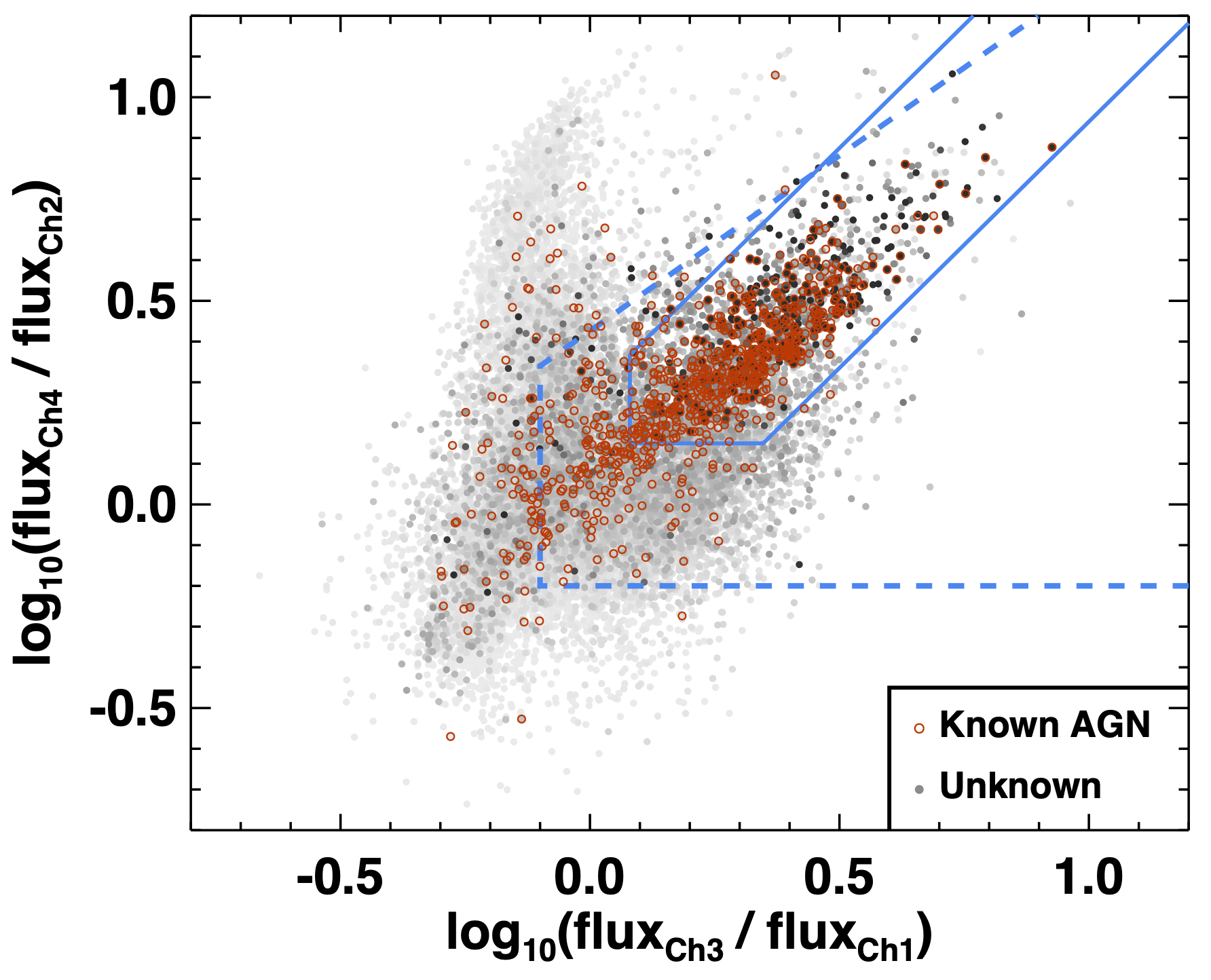}
 &
    \includegraphics[width=9cm]{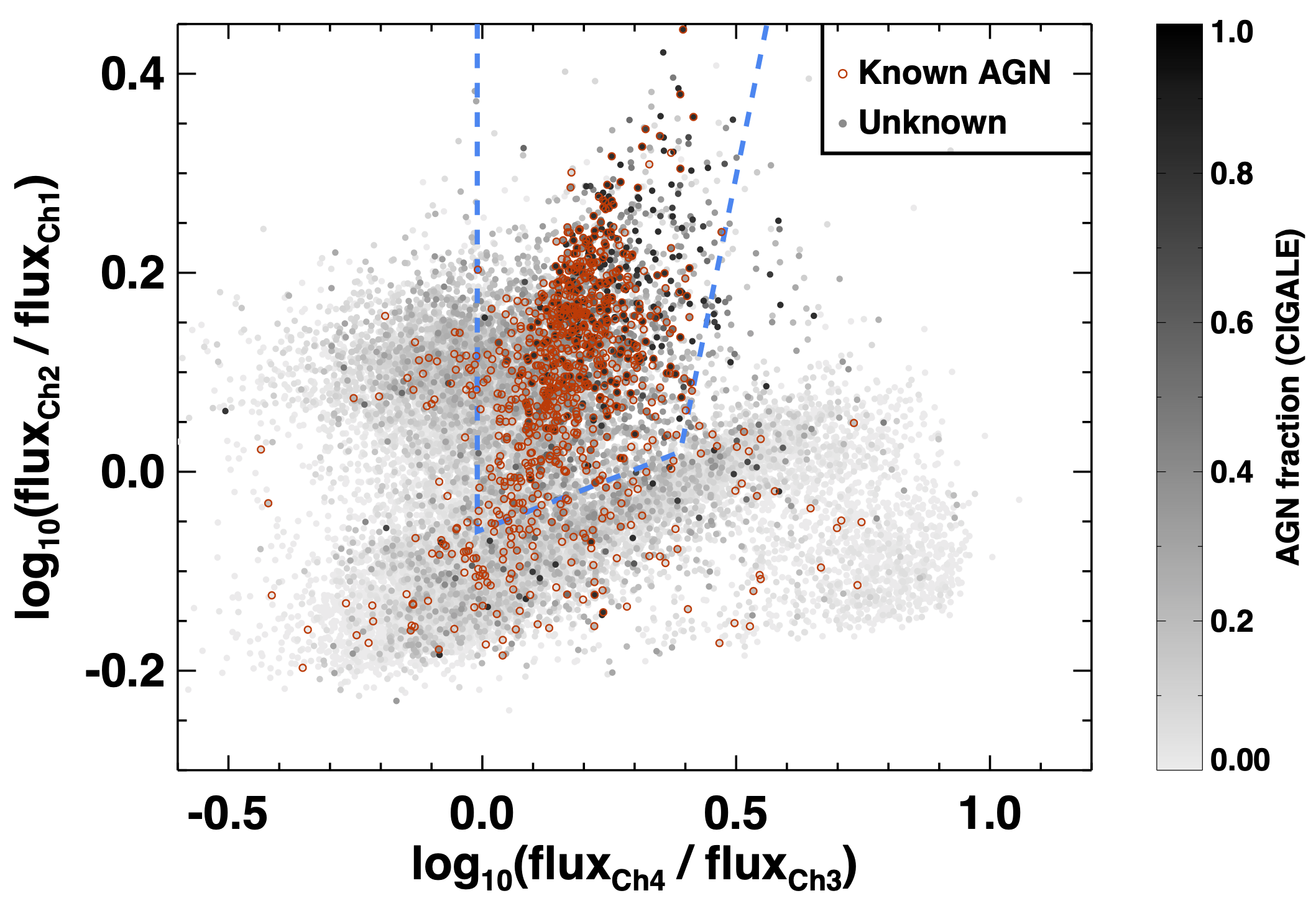}
    \end{tabular}
    \caption{The location of the LoTSS-Deep sources on the
      \citet{Lacy2004} and \citet[][left]{Donley2012} and on the
      \citet[][right]{Stern2005} mid-IR colour-colour classification
      plots (for sources with S/N$>$2 in all four bands), in the
      Bo\"otes field. The blue dashed lines on the left-hand panel
      show the Lacy et~al.\ selection criteria, and the blue solid 
      lines show those of Donley et~al. On the right-hand plot, the
      Stern wedge is shown by the blue dashed lines. In both plots,
      the greyscale colour-coding indicates the AGN fraction from the
      \cigale\ SED fitting using the \skirtor\ AGN model. Objects
      confirmed to be AGN through optical spectroscopy or X-ray
      observations are indicated by the red circles.}
    \label{fig:SDMcolourcoded}
\end{figure*}

\citet{Lacy2004} and \citet{Stern2005} derived the first colour-cuts
based on shallow {\it Spitzer} data (hereafter referred to as the Lacy
and Stern regions, respectively), and these were effective in
separating out AGN from the population of relatively nearby inactive
galaxies.  However, the broad colour regions selected in these papers
are heavily contaminated by higher redshift ($z>0.5$) inactive
galaxies, that deeper {\it Spitzer} surveys (such as those available
in the LoTSS Deep Fields) are able to detect. \citet{Donley2012}
therefore defined a much tighter region of mid-IR colour space
(hereafter, the Donley region) within which AGN samples display much
lower contamination, but consequently are also less complete. Even in
these deep datasets, however, fainter galaxies often lack measurements
in one or more channels, preventing any classification by the Stern,
Lacy or Donley criteria. To help overcome this, \citet{Messias2012}
derived a series of redshift-dependent colour cuts based on K-band to
Channel 2, Channel 2 to Channel 4, or Channel 4 to 24$\mu$m flux
ratios (hereafter, the Messias regions). These allow classification of
a larger fraction of galaxies, but with the same issues regarding
completeness and contamination. Furthermore, simple application of
colour cuts takes no account of low signal-to-noise measurements which
can scatter data across the colour criteria, and can also miss some
types of AGN \citep[e.g.][]{Gurkan2014}.

\begin{figure*}
    \centering
    \includegraphics[width=10.5cm]{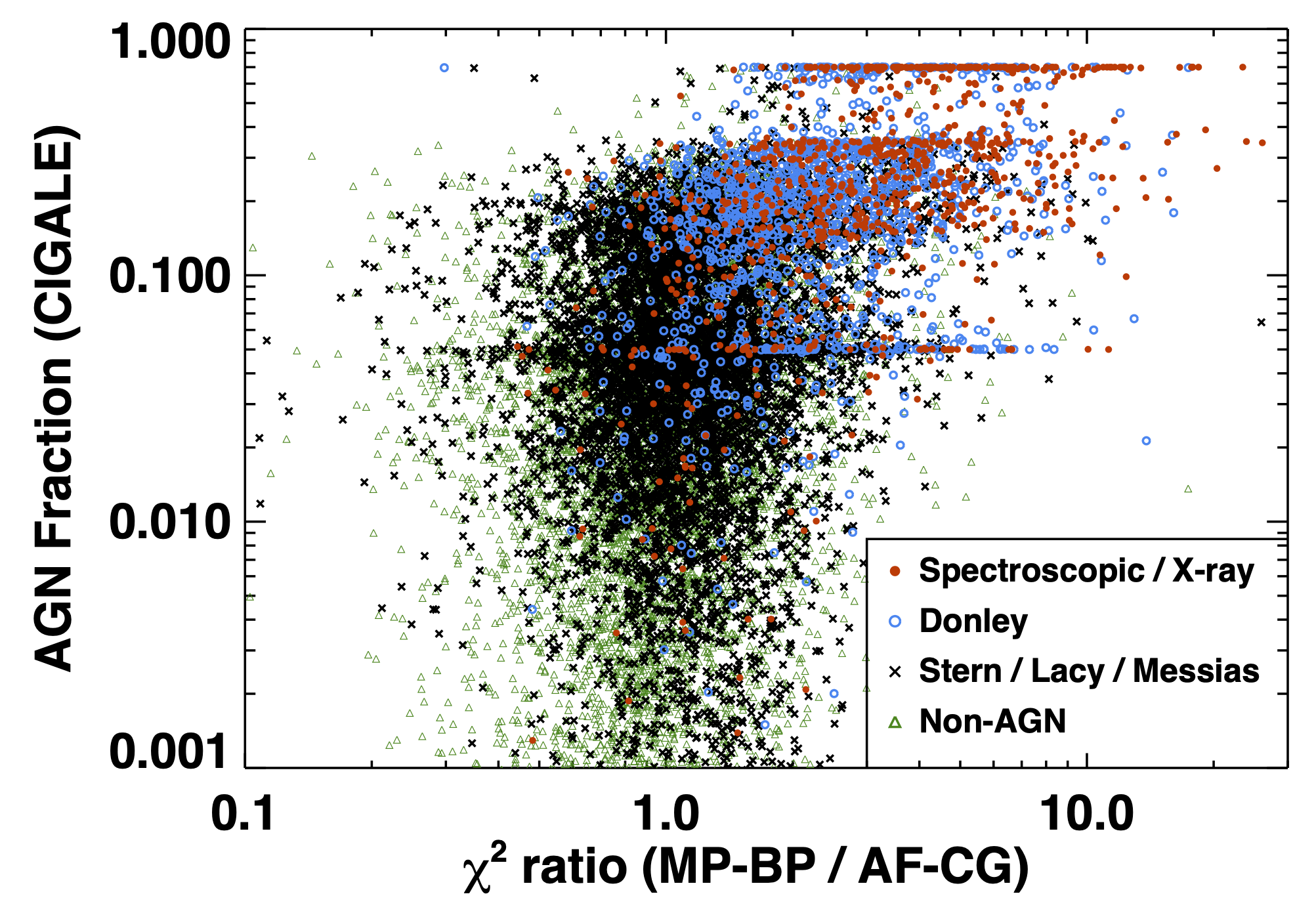}
    \caption{The \cigale\ \skirtor\ AGN fraction plotted against the
      ratio of the $\chi^2$ values between SED codes that do not
      include AGN components (the lower value for the \magphys\ (MP)
      and \bagpipes\ (BP) fits) and those that do (the lowest of the
      \cigale\ (CG) and \agnfitter\ (AF) fits), for the LoTSS-Deep
      sources in Bo\"otes. Points are colour-coded according to
      whether they are spectroscopic or X-ray AGN (red filled
      circles), or satisfy the Donley criteria (with S/N$>$3 in each
      band; blue open circles), or satisfy the broader Stern, Lacy or
      Messias cuts (with S/N$>$3; black crosses), or `non-AGN' that
      either do not satisfy any cuts or have too low signal-to-noise
      in the mid-IR for this to be determined (green triangles). The
      clustering at certain \cigale\ AGN fraction values (e.g.\ 0.05,
      0.7) appears to be a feature of the code, perhaps due to the
      fairly limited sampling of the grid of AGN model parameters. The
      plot shows that as the \cigale\ AGN fraction rises above
      $\sim$0.1, objects are more likely to be identified as AGN
      through spectroscopic or X-ray selection or the Donley mid-IR
      cuts, and also that the SED fitting begins to deteriorate
      (higher relative $\chi^2$) for SED codes that don't include AGN
      components.}
    \label{fig:AGNfracchi}
\end{figure*}

\begin{figure*}
    \centering
    \includegraphics[width=0.85\textwidth]{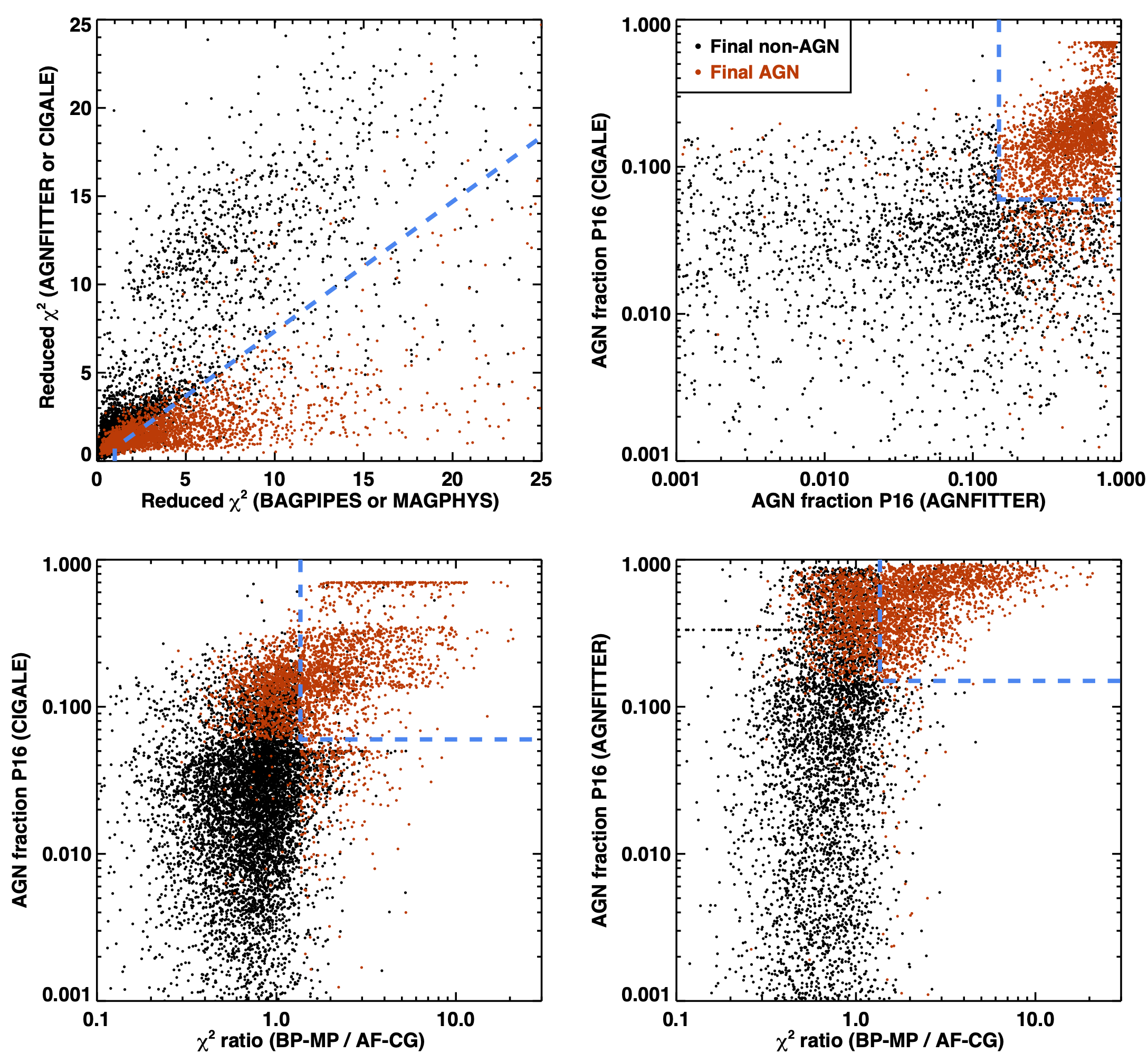}
    \caption{The selection criteria used to identify radiative-mode
      AGN, and the relative distributions of the AGN and non-AGN thus
      identified. The upper-left panel compares the reduced $\chi^2$
      value resulting from SED models including an AGN component
      (\agnfitter, \cigale) against those which do not (\magphys,
      \bagpipes), with the blue dashed line showing selection
      criterion (iii). The upper right plot shows the 1-sigma lower
      limits (16$^{\rm th}$ percentile; P16) to the AGN fraction from
      \agnfitter\ and \cigale\ (with the \skirtor\ AGN model), with
      the blue dashed lines showing selection criteria (i) and
      (ii). The lower plots show selection criteria (i) {\it vs} (ii)
      and (i) {\it vs} (iii) in the left and right panels
      respectively. Data shown are for ELAIS-N1. Sources are selected
      as radiative-mode AGN if they satisfy at least two of the three
      criteria (or are confirmed AGN from spectroscopic or X-ray
      observations): these sources are shown in red.}
    \label{fig:selectedAGN}
\end{figure*}

\begin{figure*}
    \centering
    \includegraphics[width=\textwidth]{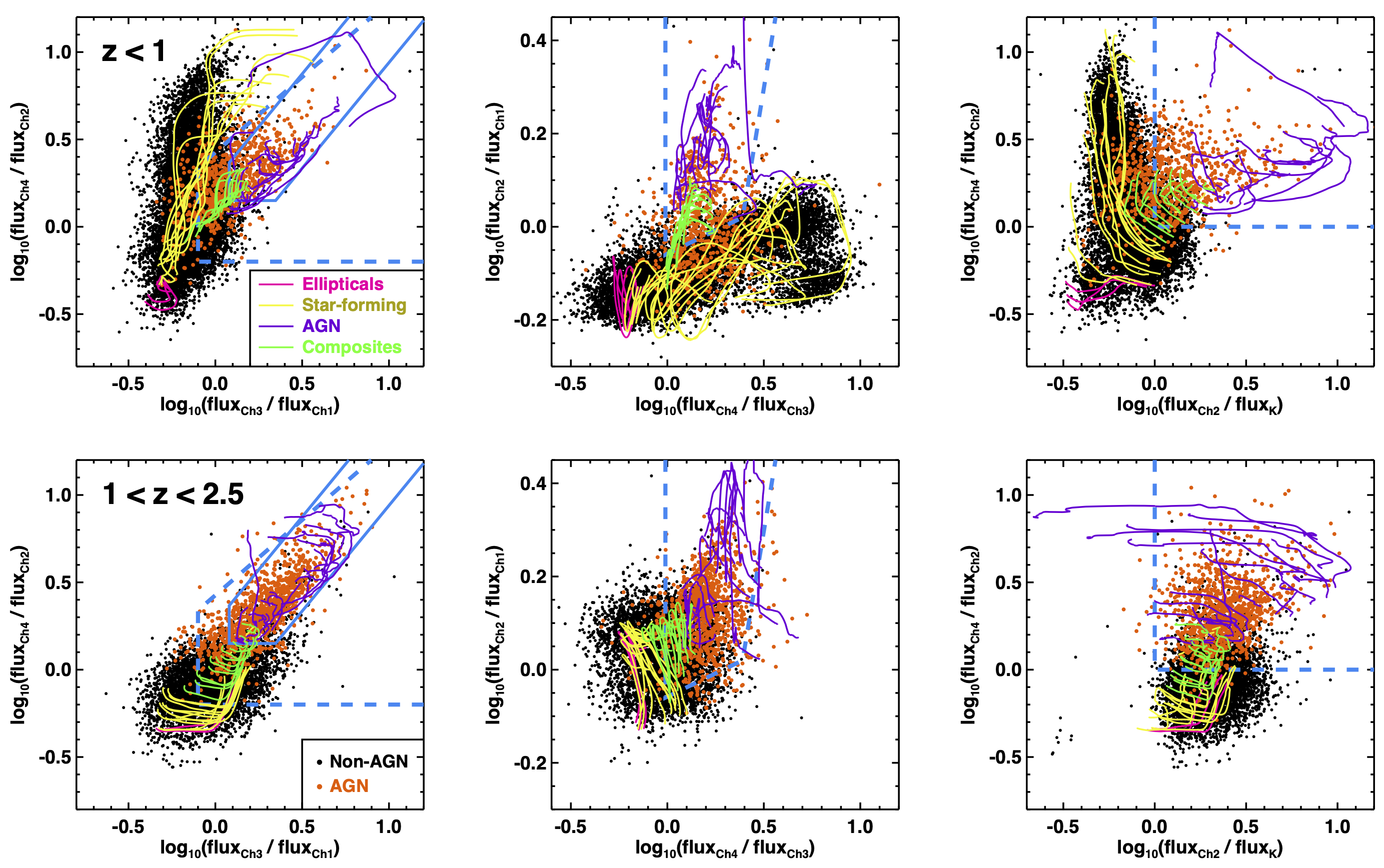}
    \includegraphics[width=\textwidth]{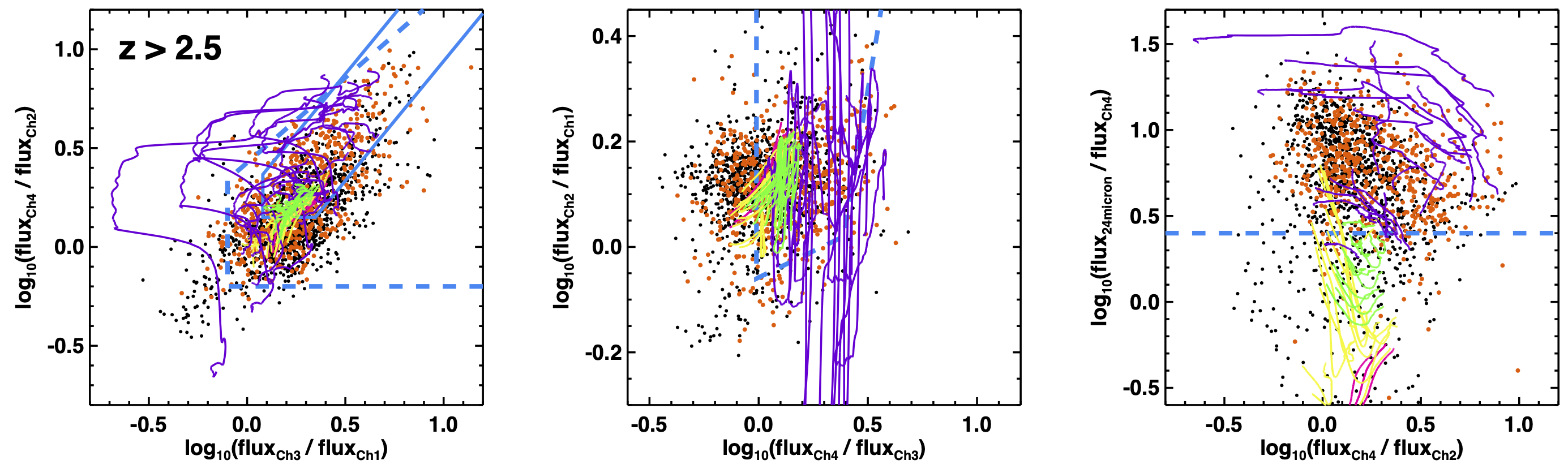}
    \caption{Infrared colour-colour plots for the LoTSS-Deep sources
      in ELAIS-N1, compared with template spectra. The sources
      classified as (radiative-mode) AGN are plotted in red and the
      non-AGN in black symbols; sources are only plotted if they have
      a signal-to-noise of at least 3 in each of the relevant
      filters. For clarity, sources (and templates) are divided into
      three redshift ranges: the top row is for $z < 1$, the middle
      row for $1<z<2.5$ and the bottom row for $z>2.5$. For each
      redshift, the left-hand plot shows the mid-IR IRAC flux ratios
      used for the \citet[][blue dashed lines]{Lacy2004} and
      \citet[][blue solid lines]{Donley2012} selections. The middle
      column shows the \citet{Stern2005} colour criteria, with the
      Stern region indicated by the blue dashed lines. The right-hand
      column shows the selection criteria proposed by
      \citet{Messias2012}, combining IRAC colours with the K-band flux
      at the lower redshifts, and with the 24$\mu$m flux at the
      highest redshifts. In each plot the coloured lines indicate the
      evolution over the specified redshift range of a selection of
      galaxy and AGN template spectra, from \citet{Brown2014} and
      \citet{Brown2019}, separated into ellipticals (pink),
      star-forming galaxies (yellow), AGN (purple), and composites
      (green). As can be seen, the broad colour cuts suffer to various
      extents from both incompletensss and contamination. The selected
      AGN broadly align with the regions of colour space covered by
      the AGN and composite template spectra.}
    \label{fig:SDMtemplates}
\end{figure*}

The wide array of data available in the LoTSS Deep Fields allows a
classification scheme to be developed which uses much more than just
the mid-IR colour bands. The SED fitting described in the previous
section encodes all of the mid-IR spectral expectations used in the
Stern, Lacy, Donley and Messias colour criteria, but combines this
with additional near-IR and optical data which allow simultaneous
characterisation of the host galaxy properties; the latter allows the
contribution of the host galaxy to the mid-IR to be directly
predicted, and thus any additional AGN contribution to be more clearly
distinguished. As an indication of this,
Figure~\ref{fig:SDMcolourcoded} shows the Stern, Lacy and Donley
mid-IR colour-colour plots with the LoTSS-Deep sources in
Bo\"otes\footnote{In Figs.~\ref{fig:SDMcolourcoded}
and~\ref{fig:AGNfracchi} the Bo\"otes field is used to show the
results, as the superior spectroscopy and X-ray coverage in this field
gives a higher quantity of `known AGN' to demonstrate the results. In
Figs.~\ref{fig:selectedAGN} to ~\ref{fig:SFRradio}, ELAIS-N1 is used
to demonstrate the results, as this is the deepest field with the best
multi-wavelength data. In all cases, all three deep fields show
consistent results.}  colour-coded by their AGN fraction as derived by
\cigale\ using the \skirtor\ model. Sources classified as an AGN
through optical spectra or X-ray properties are indicated in red. It
can be seen that the X-ray and spectroscopically selected AGN and the
objects with high \cigale\ AGN fractions concentrate primarily in the
selected colour-space regions, especially the Donley region, but that
a significant fraction of these probable AGN are also found outside of
these regions. Furthermore, there are objects within the colour-cuts
(especially the broader Lacy and Stern regions) for which
\cigale\ predicts very low AGN contributions to the mid-IR.

The use of the four SED fitting routines provides two routes to
identifying the probable AGN. First, each of \cigale\ and
\agnfitter\ provides an estimate of $f_{\rm AGN}$, the fractional AGN
contribution to the mid-IR. Second, objects which have a significant
AGN contribution to their SED should be poorly fitted using
\magphys\ or \bagpipes\ (and typically better fitted using \cigale\ or
\agnfitter). Figure~\ref{fig:AGNfracchi} demonstrates these effects,
by showing the \cigale\ AGN fraction plotted against the ratio of the
$\chi^2$ values determined from the SED fits without AGN components
compared to those with AGN components, with points colour-coded by
evidence for AGN from either spectroscopic or X-ray data, or from
mid-IR colour cuts. The spectroscopic and X-ray selected AGN generally
show both moderate-to-high AGN fractions and a higher $\chi^2$ using
\magphys/\bagpipes\ than using \cigale/\agnfitter. The majority of
objects which lie securely within the Donley mid-IR colour-cuts show
the same characteristics. Objects that lie only within the broader
Stern, Lacy or Messias colour regions typically show much lower AGN
fractions and the $\chi^2$ value from the \magphys/\bagpipes\ fits is
lower than or comparable to that from \cigale/\agnfitter; they largely
overlap with the `non-AGN' that either lie outside of these colour
cuts or do not have sufficiently high signal-to-noise in their mid-IR
measurements for this to be determined. Nevertheless, the SED fits are
able to pick out promising AGN candidates within these categories.

An examination of the AGN fractions derived by \cigale\ and especially
by \agnfitter\ shows that many of these have quite large
uncertainties, especially for fainter galaxies with fewer
securely-measured photometric points. Investigations indicated that
the 16$^{\rm th}$ percentile of the posterior of the AGN fraction
(i.e. the 1-sigma lower limit on the AGN fraction; hereafter P16)
provided a more robust indication of the presence of an AGN. The
selection of radiative-mode AGN was therefore made by considering
three selection criteria (see below for a discussion of
  how the threshold values were set):

\begin{enumerate}
\item whether the P16 AGN fraction from \cigale, using the
  \skirtor\ AGN models, exceeded a threshold value of 0.06 (ELAIS-N1
  and Lockman Hole fields) or 0.10 (Bo\"otes field).
\item whether the P16 value for the AGN fraction from \agnfitter, as
  defined in Eq.~\ref{eqn:agnfitter}, exceeded a threshold value of
  0.15 (ELAIS-N1 and Lockman Hole fields) or 0.25 (Bo\"otes field).
\item if the lower of the reduced $\chi^2$ values arising from the
  \magphys\ and \bagpipes\ SED fits was both greater than unity and at
  least a factor $f$ greater than the lowest of the reduced $\chi^2$
  values arising from the two \cigale\ and the \agnfitter\ SED
  fits. The factor $f$ was determined to be twice the median value of
  the $\chi^2$ ratio between the better fit from \magphys\ and
  \bagpipes\ and the best fit from \cigale\ and \agnfitter
  (cf.\ Figure~\ref{fig:selectedAGN}). This evaluated to $f=1.36$ for
  ELAIS-N1, $f=1.59$ for Lockman Hole and $f=2.22$ for Bo\"otes.
\end{enumerate}

\noindent An object was classified as a radiative-mode AGN if it
satisfied at least two of these three criteria. In practice, this
means either that it has a determined high AGN fraction from both
\cigale\ and \agnfitter\, or it has a high AGN fraction from at least
one of the two codes combined with a superior SED fit using methods
which include AGN components. The selection cuts for each criterion
were set by comparing the derived classifications with the
spectroscopic and X-ray samples and considering the locations of the
classified AGN and non-AGN on mid-IR colour-colour diagrams. The
threshold values selected were different for Bo\"otes than for the
other two fields. This is because the AGN fractions calculated in that
field were systematically higher than those in ELAIS-N1 or Lockman
Hole (e.g. a median AGN fraction of 0.037 in Bo\"otes using the
\cigale\ \skirtor\ model, compared to 0.029 in each of ELAIS-N1 and
Lockman), which is likely to be due to the different manner in which
the photometric catalogues were constructed in Bo\"otes
\citepalias[see][]{Kondapally2021}. Setting higher thresholds in
Bo\"otes ensured a consistency of classification across the three
fields (cf. Sec.~\ref{sec:classes}). Finally, a small proportion of
objects did not meet these criteria but had previously been identified
to be an AGN based on either optical spectra or X-ray properties;
these were added to the radiative-mode AGN sample (and correspond to
about 3 per cent of all radiative-mode AGN).

Fig.~\ref{fig:selectedAGN} shows the LoTSS-Deep sources on different
combinations of these selection criteria, with the sources that
satisfy at least two criteria, and therefore are selected as
radiative-mode AGN, shown in red. It can be seen that there is a broad
consistency between the different criteria: most of the selected
radiative-mode AGN satisfy all three criteria and therefore are secure
classifications. The main addition to this is a population of sources
selected as having high AGN fractions by both \cigale\ and
\agnfitter\ but with comparable, low $\chi^2$ values from the
different fitting methods; these are probably sources where
\cigale\ and \agnfitter\ are able to pick out a weak AGN through the
mid-IR emission, but there is little-to-no direct AGN light through
the optical to near-IR spectrum and so \magphys\ and \bagpipes\ are
still able to provide a good fit to the majority of the spectrum.

Fig.~\ref{fig:SDMtemplates} shows the selected radiative-mode AGN and
non-AGN on a series of mid-IR colour-colour diagrams, compared against
the evolving colours of various galaxy template models. The panels are
split by redshift ranges, in order to allow a clearer comparison
against the template expectations. At each redshift, the panels show
the Lacy and Donley colour plots (left), the Stern colour plot
(middle), and the appropriate Messias plot (right). Template SED
models were drawn from the `Galaxy SED Atlas' of \citet{Brown2014}
combined with the `AGN SED Atlas' of \citet{Brown2019}. SEDs were
selected from these libraries for: (i) elliptical galaxies (as
expected to be seen for jet-mode AGN); (ii) star-forming galaxies;
(iii) AGN (including both quasars and edge-on `type-II' AGN); and (iv)
composite spectra, produced by combining a set of Seyfert AGN spectra
with host galaxy spectra, with a range of weights.

The template tracks for the different galaxy classes confirm both the
motivation for, and the shortcomings of, the colour-colour selection
criteria: the Donley region relatively cleanly selects AGN at $z<2.5$
but is incomplete for composite systems; the Stern and Lacy regions
are more complete for composite systems but contaminated, especially
at the higher redshifts; the Messias cuts perform relatively well,
especially at the highest redshift where the use of the 24$\mu$m
colour gives a clear advantage, but still have some incompleteness and
contamination. The red points show the objects selected as
radiative-mode AGN by the techniques outlined above. At all redshifts
these broadly overlap the regions of the AGN and composite templates,
extending where appropriate beyond the colour-selection limits. It is
clear, however, that in the $z>2.5$ redshift range there remains a
significant population of objects that are not classified as AGN, and
yet which lie in similar regions of colour-space to the AGN. At these
redshifts, as is evident from Fig.~\ref{fig:SDMtemplates}, it is only
the Channel 4 and 24$\mu$m filters that are able to probe rest-frame
wavelengths where an AGN template becomes clearly distinct from the
galaxy templates, and the composites are even more difficult to
distinguish. Especially with the typically low signal-to-noise of the
galaxies in this highest redshift bin, the SED fitting techniques may be
less reliable: although the classifications are provided for all
sources, readers should treat these with caution at $z>2.5$, where
there may well be a degree of incompleteness in the AGN sample.

\section{Comparison of derived properties and consensus measurements}
\label{sec:consensus}

Two of the most important galaxy properties to determine are the
stellar mass and the star-formation rate. Each of the SED fitting
codes provides an estimate of these parameters. This section discusses
how these values are combined to produce consensus measurements for
each source.

In brief summary, for sources which do not host an AGN, the
\magphys\ and \bagpipes\ codes ought to provide the best measurements
of mass and SFR, because these models offer a significantly broader
selection of galaxy templates. Indeed, for these sources, the results
from these two codes show excellent agreement in their estimates of
both stellar mass (median absolute difference of just 0.09 dex) and
SFR (0.14 dex). The consensus values of the stellar mass and SFR for
non-AGN were therefore generally derived from the logarithmic mean of
the \magphys\ and \bagpipes\ results.

For radiative-mode AGN, the \magphys\ and \bagpipes\ results are
potentially unreliable as they do not include any AGN component in
their SED modelling. The two \cigale\ runs (with the Fritz and
\skirtor\ AGN models) should be more reliable, and indeed these two
agree with each other well: the median absolute difference is only
0.09 dex in stellar mass and 0.13 dex in SFR. \agnfitter\ is found to
provide less consistent results, but is valuable for the small
fraction ($\approx 2$ per cent) of sources which are highly
AGN-dominated, and for which \agnfitter's superior modelling of the
AGN UV emission is required. The consensus values of the stellar mass
and SFR for radiative-mode AGN were therefore typically derived from
the logarithmic mean of the two \cigale\ results, except where
\cigale\ failed to provide an acceptable fit, in which case the
\agnfitter\ values were adopted.

Sections~\ref{sec:conc_mass} and~\ref{sec:conc_sfr} now provide (for
stellar mass and SFR respectively) a much more detailed comparison of
the outputs of the different SED fitting codes, along with a full
description of how the generalised approach discussed above was
adapted in cases where one or more of the SED codes failed to provide
an acceptable fit. Readers not interested in these finer details may
wish to skip to Section~\ref{sec:radAGN}.

\subsection{Consensus stellar masses}
\label{sec:conc_mass}

For sources which are not identified to be a radiative-mode AGN, the
results from the \magphys\ and \bagpipes\ codes show excellent
agreement in their estimates of stellar mass: where both \magphys\ and
\bagpipes\ pass the threshold for an acceptable fit (see
Section~\ref{sec:seds}) the median absolute difference in stellar mass
is just 0.09 dex, with over 90 per cent of sources agreeing within
0.25 dex; the outliers are generally the faintest sources, at low
masses or high redshifts. \cigale\ also gives very similar values,
with a median difference in stellar mass of only 0.11 dex, and over 85
per cent agreeing within 0.25 dex. \agnfitter\ shows much lower
agreement, however, with a median difference in stellar mass of 0.27
dex compared to the estimates from the other codes. This inconsistency
for \agnfitter\ is likely to be associated with the lack of an energy
balance in the fitting process.

\begin{figure*}
    \centering
    \includegraphics[width=0.85\textwidth]{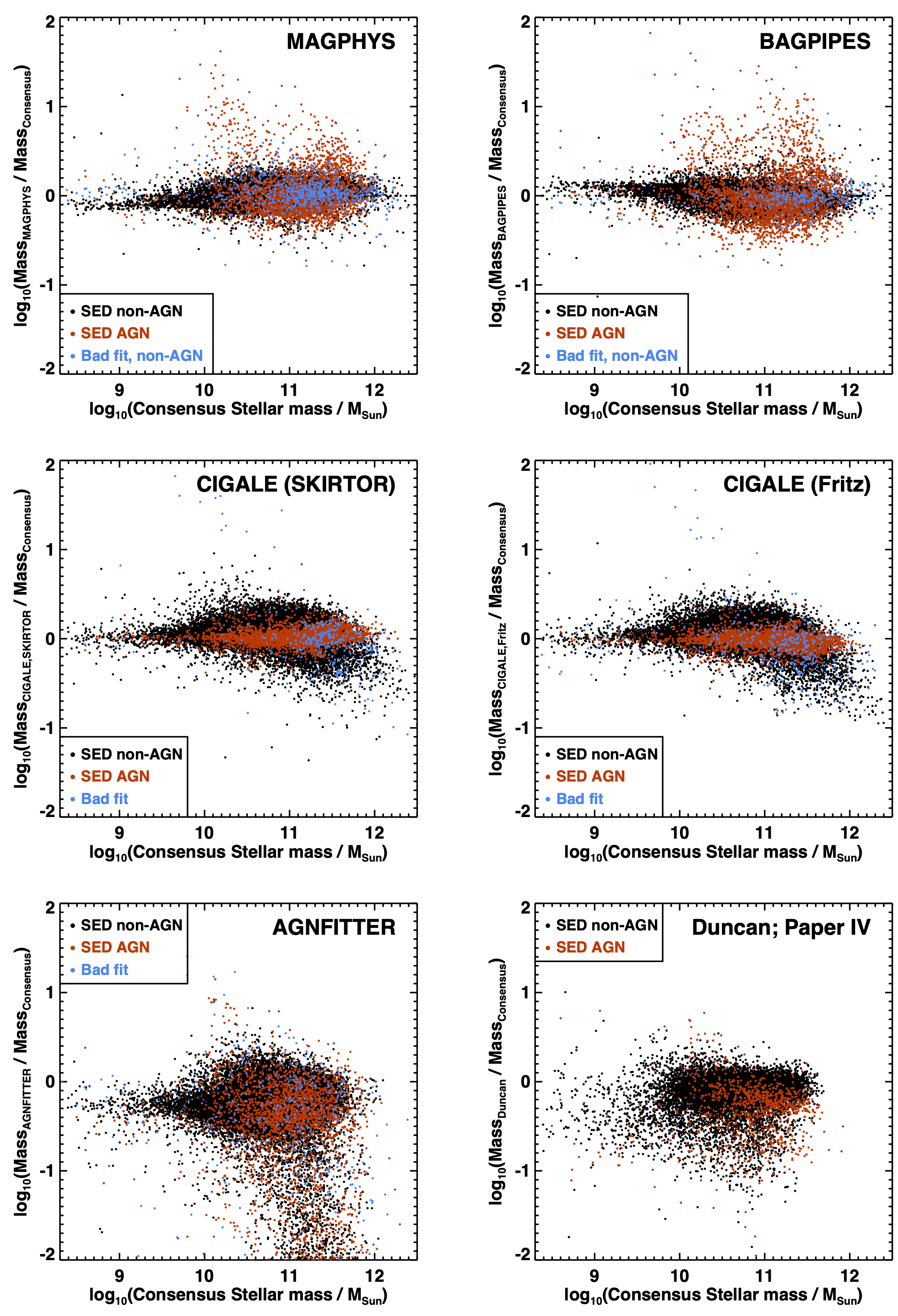}
    \caption{A comparison of the masses derived by the different SED
      fitting codes against the final consensus masses for the LoTSS
      Deep Field sources in ELAIS-N1. \magphys, \bagpipes\ and
      \cigale\ all give broadly consistent results for non-AGN, but
      differ for the AGN subset, for which the \cigale\ results should
      be more reliable. \agnfitter\ masses show a small systematic
      offset compared to the other codes, and more outliers at high
      mass. The lower right plot examines the masses produced in
      \citetalias{Duncan2021} (only out to $z<1.5$); these are seen
      to give consistent results with only slightly larger scatter.
      This is of interest because these stellar
      masses were produced for the entire galaxy population in these
      deep fields, not just the LoTSS-Deep host galaxies.
    }
    \label{fig:mass_consensus}
\end{figure*}

For these non-AGN the consensus stellar mass was derived from the mean
of the logarithm of the stellar masses derived using \magphys\ and
\bagpipes, as long as both codes provided an acceptable fit to the
data ($\approx 86$ per cent of the non-AGN, though rising to nearly 95
per cent in ELAIS-N1).  If one of the two codes provided a bad fit and
the other a good fit (11 per cent of cases), then the stellar mass
estimate from the well-fitting code was adopted as the consensus
measurement. If both codes produced fits below the acceptability
threshold then the values of the two stellar mass estimates were
examined: if they agreed with each other within 0.3 dex ($\approx 2$
per cent of cases) then it was likely that the unreliability of the
SED fits was driven by some outlier points that did not invalidate the
stellar mass estimates, and so the logarithmic mean of the two values
was adopted as the consensus stellar mass. If the two values disagreed
by more than 0.3 dex, then the stellar mass estimates of the two
\cigale\ fits were examined as well: if the full range of all 4
stellar masses was less than 0.6 dex ($\approx 0.3$ per cent of cases)
then the logarithmic mean of the four measurements was adopted as the
consensus measurement; if the range was larger than 0.6 dex ($\approx
0.6$ per cent of sources) then it was deemed that no reliable stellar
mass could be provided. A comparison of the consensus masses derived
against the estimates from each code individually is shown by the
black points in Fig.~\ref{fig:mass_consensus}, confirming visually the
good agreement of the \magphys\ and \bagpipes\ codes, broad agreement
of \cigale, and larger scatter of \agnfitter\ for these sources.

For radiative-mode AGN, the two \cigale\ runs provide stellar mass
estimates that agree well with each other: the median absolute
difference is only 0.09 dex, with 90 per cent of sources within 0.3
dex. Compared to these values, as expected, the results from
\magphys\ and \bagpipes\ show greater scatter (each 0.16 dex median
difference) and also a larger fraction of outliers where the codes
significantly over-estimate the mass due to AGN light being
incorrectly modelled as stellar emission
(cf. Fig.~\ref{fig:mass_consensus}).  Again, \agnfitter\ shows a
larger dispersion in stellar mass measurements relative to the other
codes, with a median absolute difference of 0.49 dex; this may be due
to the stellar component being fitted independently without an energy
balance constraint, with some stellar light perhaps being incorrectly
modelled as AGN emission or vice versa, although it could also be
related to the different approach to modelling the AGN emission. For
these reasons, for the radiative-mode AGN, if both \cigale\ runs
provided acceptable fits then the logarithmic mean of the stellar
masses from these two runs was accepted as the consensus mass (with
\agnfitter\ excluded due to its higher proportion of outliers); this
was the case for just over 94 per cent of the radiative-mode
AGN. Otherwise, if just one of the \cigale\ runs provided an
acceptable fit ($\approx 3$ per cent of cases) then the stellar mass
from that run was adopted. If neither \cigale\ run provided a good
fit, but \agnfitter\ did, then there was a likelihood that this was a
case where either energy balance was breaking down or the superior
modelling of the AGN UV emission by \agnfitter\ was helping the fit;
in these 2 per cent of cases, the \agnfitter\ stellar mass estimate
was used. Otherwise, it was decided that no reliable stellar mass
estimate was possible.

Fig.~\ref{fig:mass_consensus} shows a comparison on each mass estimate
against the consensus mass derived, and illustrates the trends
discussed above. The lower-right panel also compares the consensus
masses against those derived in \citetalias{Duncan2021} using a
grid-based SED fitting mechanism \citep[see also][]{Duncan2019}. This
comparison is interesting because the stellar masses in
\citetalias{Duncan2021} are derived for all galaxies in the field, not
only the radio sources, and therefore allow a comparison between the
radio sources and the underlying population. In
\citetalias{Duncan2021} it is argued that the stellar mass estimates
are only reliable out to $z \sim 1.5$, and so this is set as an upper
limit for the plotted points. As can be seen, the agreement between
the \citetalias{Duncan2021} stellar masses and the consensus masses
derived here is very good for the non-AGN, with no significant
systematic offset ($<0.1$ dex) and a median scatter of 0.11 dex. The
performance for AGN is slightly worse, but still good, with a median
scatter of 0.23 dex. These results confirm that the
\citetalias{Duncan2021} masses provide reliable measurements for the
broader population that can be used in comparison against the
consensus masses for the radio source population.

In this paper, no attempt is made to derive uncertainties on the
consensus stellar masses for individual sources. Uncertainties arise
both due to statistical errors in the individual fits and systematic
effects between different SED codes. Each SED code offers an estimate
of its statistical uncertainty for each source, and the difference
between the stellar masses from different SED codes can be used to
gauge the size of the systematic errors. Another source of error is
that during the SED fitting the redshift of the source is fixed at the
best photometric redshift (unless a spectroscopic redshift is
available): uncertainties in the photometric redshift are likely to be
a significant contributor to the mass uncertainty for any given
source. Instead of calculating uncertainties for individual sources,
therefore, the approach taken here is to derive characteristic
uncertainties on stellar mass as a function of the galaxy's mass and
redshift. The characteristic uncertainties are evaluated in
Appendix~\ref{app:masserrors}, and are found to be typically around
0.1 dex for higher mass sources at $z < 2$, increasing towards higher
redshifts and lower masses.

\subsection{Consensus SFRs}
\label{sec:conc_sfr}

\begin{figure*}
    \centering
    \includegraphics[width=0.83\textwidth]{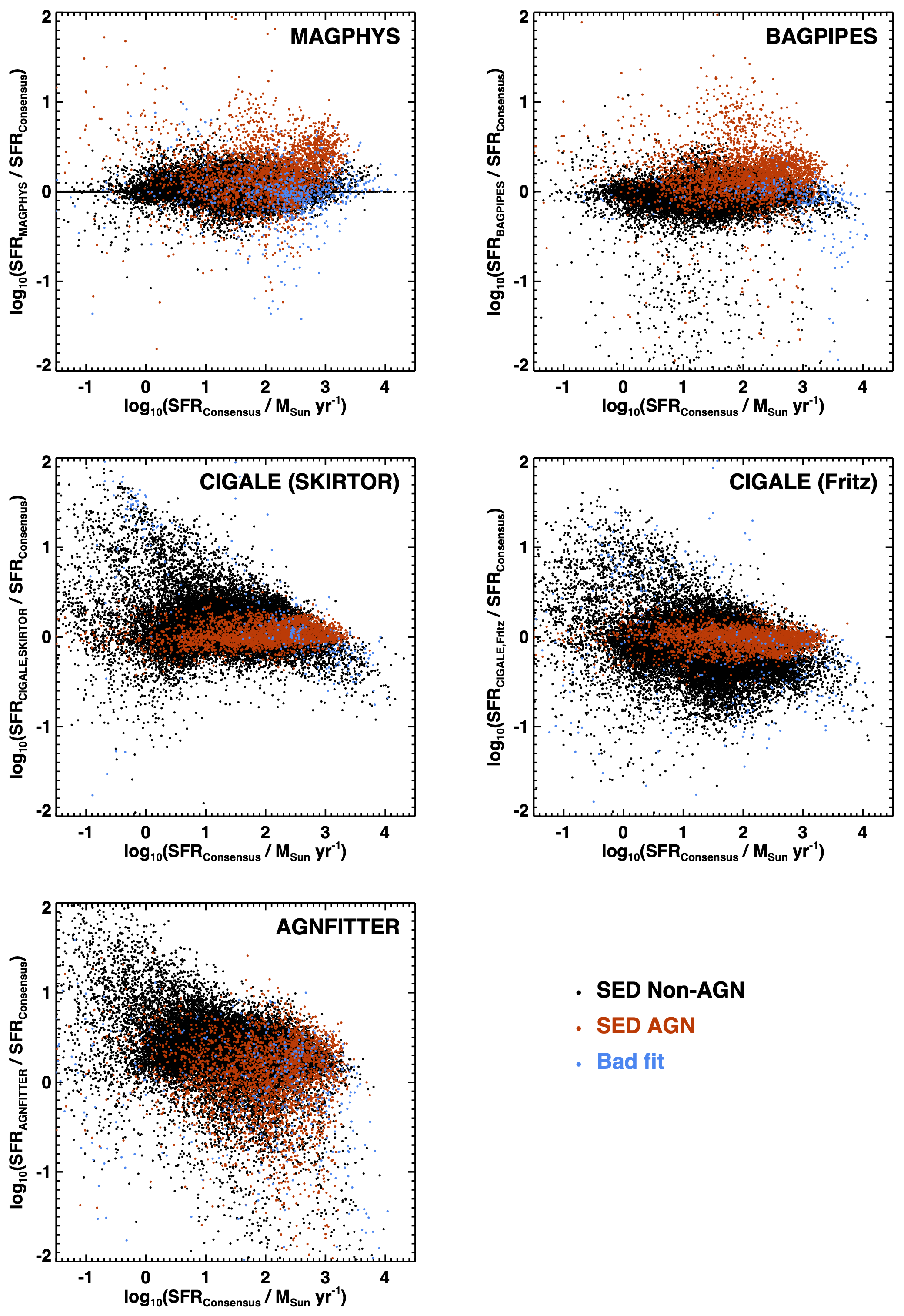}
    \caption{A comparison of the star-formation rates derived by the
      different SED fitting codes against the final consensus value,
      for the LoTSS Deep Field sources in ELAIS-N1. \magphys\ and
      \bagpipes\ give broadly consistent results for non-AGN; their
      performance on objects identified as (radiative-mode) AGN is
      more mixed, but generally reasonable where the fit is not
      flagged as a bad fit. \cigale's SFR estimations for non-AGN
      generally perform well at higher SFRs (especially with the
      \skirtor\ AGN model), but over-predict the SFR in some lower-SFR
      galaxies. The estimated SFRs of objects selected as AGN show a
      high degree of consistency between the two different
      \cigale\ runs. \agnfitter\ SFRs show more scatter and a small
      systematic offset compared to the other codes.}
    \label{fig:SFR_consensus}
\end{figure*}

Estimation of consensus SFRs follows broadly the same principles as
those of the stellar masses, in the preferred use of the \magphys\ and
\bagpipes\ results for the non-AGN and with the \cigale\ results
generally used for the AGN. As would be expected
\citep[cf.][]{Pacifici2022}, the agreement in SFR estimates between
the different codes is not quite as good as that of stellar masses,
but still strong. For non-AGN, the SFR estimates of \magphys\ and
\bagpipes\ show systematic differences of less than 0.1 dex, with a
median scatter of only 0.14 dex and over 75 per cent of cases agreeing
within 0.3 dex. The \cigale\ measurements agree comparably well at
large SFRs, but frequently provide higher SFR estimates than either
\bagpipes\ or \magphys\ at lower SFRs. \agnfitter\ suffers from a
significant systematic offset of, on average, more than 0.3 dex higher
SFRs than the other estimators. For the radiative-mode AGN, the two
\cigale\ SFR estimations show good agreement with each other (median
difference 0.13 dex). Both \magphys\ and \bagpipes\ systematically
over-estimate the SFRs of these radiative-mode AGN, by around 0.15 dex
on average. Fig.~\ref{fig:SFR_consensus} provides a visual
illustration of these effects.

To determine the consensus SFRs, like for stellar masses, the outputs
from \magphys\ and \bagpipes\ are primarily considered for the
non-AGN. The only significant difference in approach arises because of
a small proportion of sources (around 9 per cent of all the non-AGN
sources, mostly at lower SFRs) for which \bagpipes\ returns an
acceptable fit, but the SFR is dramatically below that of
\magphys\ and with an uncertainty that can be several orders of
magnitude larger than the estimated value. These very low SFRs arise
because of the parametric (exponentially-declining) form of the
\bagpipes\ SFR history, which can lead to unrealistically-low best-fit
SFRs at large ages where the e-folding time is short, but with
considerable uncertainty. For these sources, the \cigale\ SFR
estimates are found to broadly agree with the \magphys\ values, with
both often within the 1$\sigma$ confidence interval of the
\bagpipes\ fit. Therefore, sources for which the \bagpipes\ fit is
deemed to be good, but the uncertainty on the \bagpipes\ SFR estimate
is more than 5 times the estimate itself, are treated differently. In
these cases, if \magphys\ provides an acceptable fit then the
\magphys\ estimate is adopted as the consensus value; if it does not,
but the \magphys\ and \cigale\ estimates agree within 0.5 dex then the
logarithmic mean of the \magphys\ and \cigale\ values is taken as the
consensus value; otherwise, the results are deemed inconsistent and no
consensus SFR is derived. Other than these cases, the approach to
derive consensus SFRs for the non-AGN exactly matches that for
deriving stellar masses. Similarly, for the radiative-mode AGN, the
approach for stellar masses using \cigale\ (or occasionally
\agnfitter) estimates is replicated for the SFRs.

Fig.~\ref{fig:SFR_consensus} compares the consensus SFRs against the
estimates from each individual code. The spread in
  derived values between different codes is comparable to that in the
  analysis of \citet{Pacifici2022}. As with stellar masses, no
attempt is made to provide a source-by-source uncertainty on the
consensus SFR, but Appendix~\ref{app:masserrors} discusses the typical
errors; except for the few per cent of lowest-SFR objects at each
redshift (where the uncertainties increase greatly), these can be
broadly approximated as $\Delta$(SFR) $\approx 0.1 (1+z)^{0.5}$ dex.

\section{Identification of radio AGN}
\label{sec:radAGN}

As discussed in the introduction, star-forming galaxies show a tight
correlation between their radio luminosity and their
SFR\footnote{Note that this assumes that effects such as
free-free absorption at low radio frequencies are not
important. \citet{Schober2017} estimate that for star-forming galaxies
like the Milky Way, free-free absorption is only important below a
critical frequencies of a few MHz, which is well below the LOFAR
observing frequency. For starburst galaxies like Arp 220, however,
they estimate a critical frequency of a few hundred MHz; this is
potentially relevant, since the LOFAR-detected sources at $z \sim 2$
have SFRs approaching those of starburst systems, and are observed at
rest-frame frequencies of $\sim 500$\,MHz. Nevertheless,
\citet{Calistro2017} studied the radio spectral shapes of
LOFAR-selected star-forming galaxies, and \citet{An2023} recently
extended this analysis to the LoTSS Deep Fields: in both cases, a
slight flattening of the median radio spectra was found at the lowest
frequencies, from $\alpha \approx 0.8$ at high frequencies to $\alpha
\approx 0.6$ at LOFAR frequencies. Although this might be evidence
for free-free absorption, this change in spectral index only affects
the radio luminosity (and hence estimated SFR) by $\approx 0.1$\,dex
for an average source. It works in the direction of reducing any radio
excess, and thus more securely classifying a source as not having a
radio AGN. Therefore, the possible effects of free-free absorption are
ignored in this paper.}. This relation allows the
identification of sources which possess significant radio emission
associated with AGN activity, as they will appear offset to larger
radio luminosities than would be predicted from their SFR
\citep[cf.][]{Delvecchio2017,Williams2018,Whittam2022}.  Relationships
between SFR and low frequency radio luminosity have been previously
derived at relatively low redshifts by \citet{Calistro2017},
\citet{Brown2017}, \citet{Gurkan2018} and \cite{Wang2019}, and most
recently by \citet{Smith2021} using the LoTSS-Deep data in
ELAIS-N1. As discussed by \citeauthor{Smith2021}, in order to
determine an accurate relation it is essential to properly account for
non-detections, otherwise there is a risk that the derived relation
will be dependent on the depth of the radio imaging, with the bias
decreasing as the depth of the radio imaging
increases. \citeauthor{Smith2021} derive their relationship out to $z
\approx 1$ using a near-IR magnitude selected sample, finding $\log_{10}
L_{\rm 150 MHz} = 22.22 + 1.06 \log_{10}({\rm SFR})$ for the sample as a
whole (where $L_{\rm 150 MHz}$ is in units of W\,Hz$^{-1}$ and SFR in
units of $M_{\odot}$\,yr$^{-1}$), based on SFRs derived using
\magphys.

\begin{figure*}
    \centering
    \includegraphics[width=0.48\textwidth]{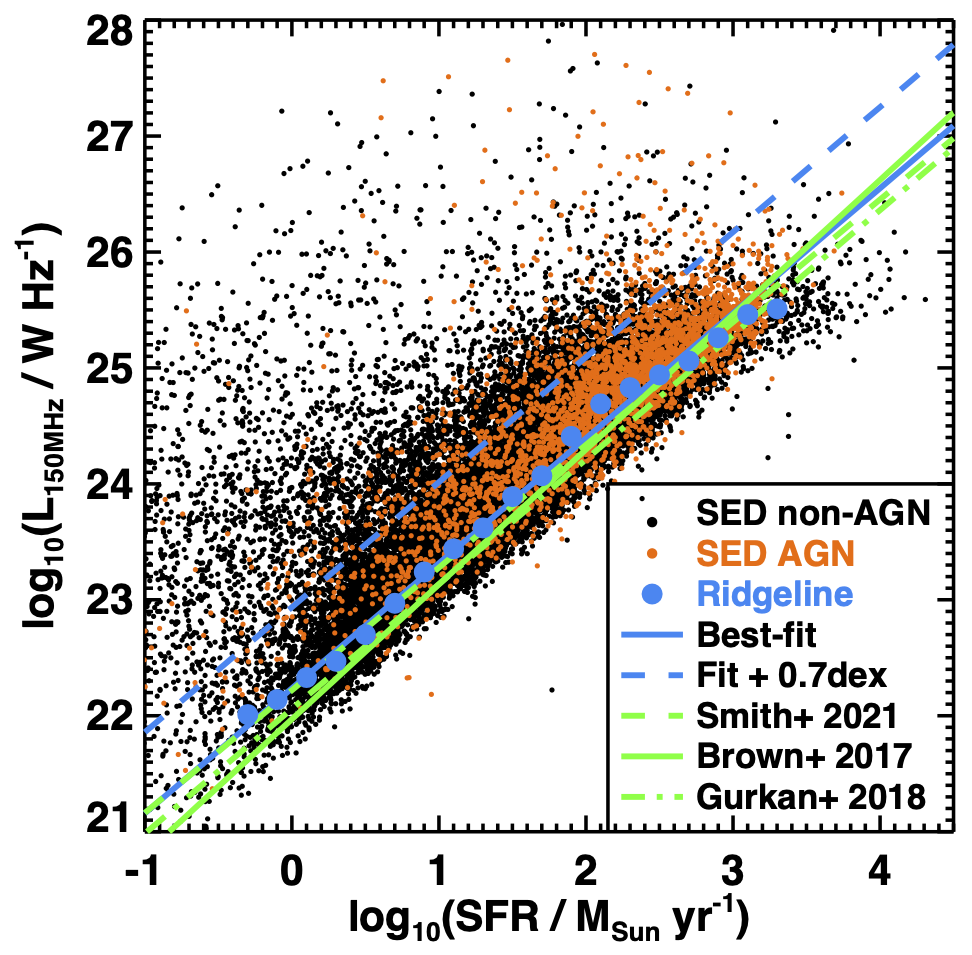} \\
    \includegraphics[width=0.9\textwidth]{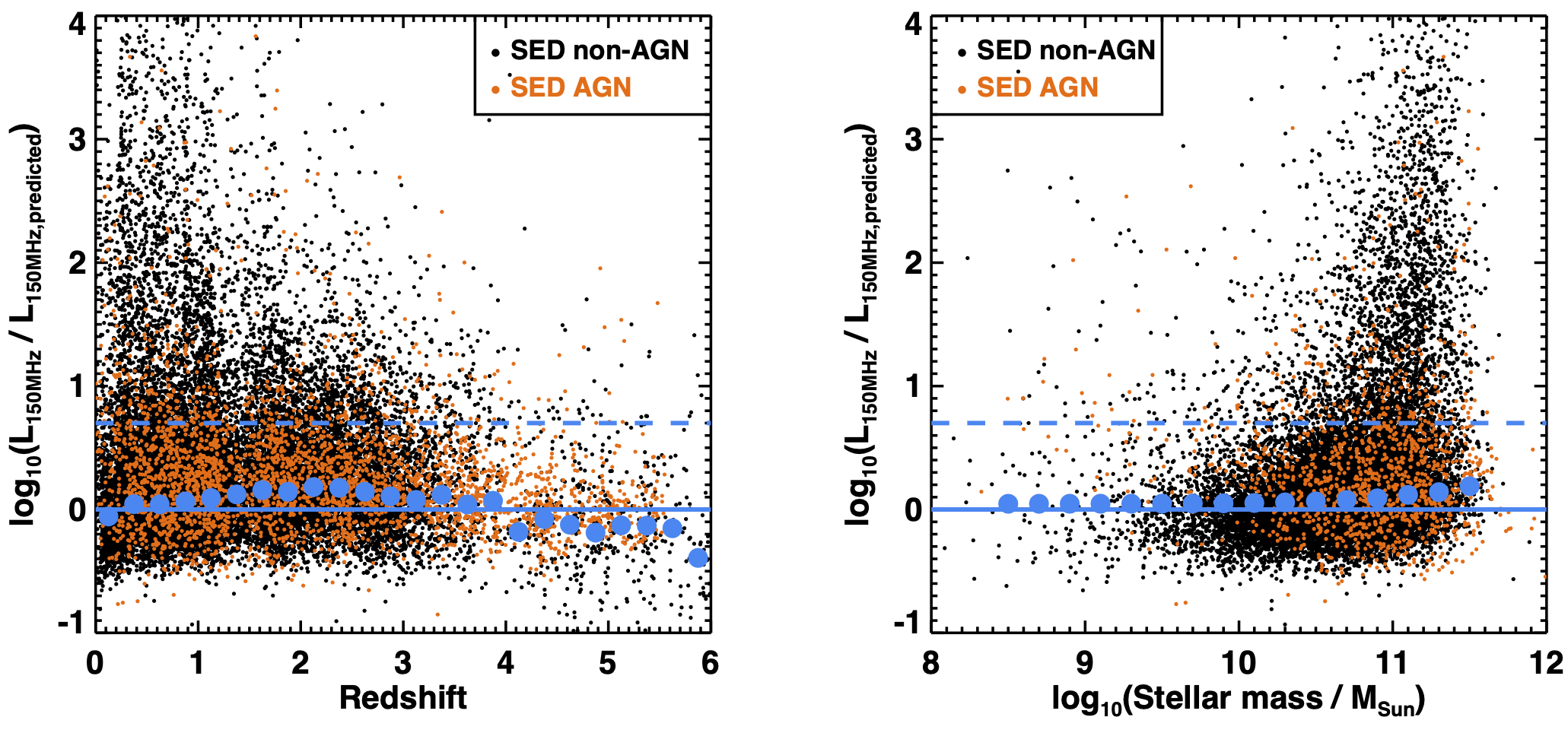}
    \caption{{\it Top:} the distribution of radio luminosity versus
      SFR for LoTSS Deep Field sources in ELAIS-N1, split into those
      identified as radiative-mode AGN from their SED (red points) and
      the sources which are not radiative-mode AGN (`SED non-AGN';
      black points). Within narrow bins in SFR, the `ridgeline' points
      (larger blue circles) indicate the peak of the distribution of
      radio luminosities. These can be well-fitted by a power-law
      distribution shown by the solid blue line, which is in broad
      agreement with literature relations (green lines). {\it Bottom:}
      the ratio of observed radio luminosity to that predicted from
      the consensus SFR based on the ridgeline fit, versus redshift
      (left) and stellar mass (right). The horizontal dashed lines
      represent the expected relation and the radio-excess
      threshold. Solid blue points in each plot show the peak of the
      distribution in narrow bins. These always lie within 0.2\,dex of
      the expected relation. Radio-excess sources are found over the
      full range of redshifts, but predominantly concentrate at high
      stellar masses.}
    \label{fig:SFRradio}
\end{figure*}

In this paper, the use of the consensus SFRs, and the extension to
higher redshifts, may be expected to lead to small changes in the
best-fit relation. A suitable relation is therefore derived using a
`ridgeline' approach. In this approach, the sources are binned into
different (narrow) bins in SFR, and within each bin the distribution
of radio luminosities of the detected sources is examined. The peak of
the distribution is identified as the ridgeline point. Provided the
radio survey is sufficiently deep then, especially in the presence of
a distorted distribution (the star-forming population plus a
distribution of radio-excess AGN), this method should provide a more
reliable value than the mean or median of the distribution of detected
sources. The radio luminosities and SFRs of the LoTSS-Deep sources are
shown in the upper panel of Fig.~\ref{fig:SFRradio}, along with the
calculated ridgeline points, which can be well-fitted by the relation

\begin{equation}
  \log_{10} (L_{\rm 150 MHz} / {\rm W\,Hz}^{-1}) = 22.24 + 1.08 \log_{10}({\rm SFR} / {\rm M}_{\odot}\,{\rm yr}^{-1})
\end{equation}

The uncertainty on the ridgeline gradient is $\pm 0.06$, and the
uncertainty on the intercept at $\log_{10}({\rm SFR})=1.5$ (the median
value, where the errors on the gradient and intercept are
uncorrelated) is $\pm 0.07$. To within 1$\sigma$, there is no
difference in this relation between those sources classified as
radiative-mode AGN or not. The relation derived from the ridgeline is
fully consistent with that of \citet{Smith2021}, agreeing within 0.1
dex over the full range of star-formation rates probed.

The distribution of radio luminosities below the ridgeline can be
reasonably well-fitted by a Gaussian distribution of width 0.22 dex;
this also holds in different bins of star-formation rate, with the
Gaussian width remaining constant (to $\pm 0.02$ dex) from low to high
SFR. The distribution above the ridgeline shows a much more extended
tail, as expected. In ELAIS-N1 and Lockman Hole, radio-excess sources
are here defined as those sources with radio luminosities exceeding
the ridgeline value by 0.7 dex, corresponding to approximately
3$\sigma$. It should be noted that this limit corresponds to
approximately 0.8 dex above the relation of \citet{Smith2021} at high
SFR; these authors derived a scatter in their relation of around 0.3
dex at ${\rm SFR} > 10 M_{\odot}$yr$^{-1}$ (at lower SFRs they
measured lower scatter, but noted that this might be due to the
limiting depth of the radio imaging); \citet{Cochrane2023} also derive
a similar value for the scatter. Therefore, the radio-excess selection
adopted here also broadly corresponds to a 3$\sigma$ excess relative
to the \citeauthor{Smith2021} relation. In Bo\"otes (where the input
photometry was different), it is found that the scatter in the
SFR-radio relation increases towards higher redshifts, and adoption of
a fixed 0.7 dex cut-off leads to an excess of radio-AGN at higher
redshifts compared to the other two fields. To remedy this, in
Bo\"otes the radio excess threshold is modified slightly to ($0.7 +
0.1z$) dex, which brings the classifications in this field in line
with those in ELAIS-N1 and Lockman (cf. Fig~\ref{fig:sourcepops}).

There is a small population of radio sources with consensus SFRs well
below $0.01 M_{\odot}$yr$^{-1}$. SFRs at this level cannot be
accurately estimated by the SED fitting codes, and thus have large
associated uncertainties. This makes a radio-excess classification
based on the consensus SFR potentially unreliable for these
sources. To avoid this issue, these sources were only classified as
radio-excess if their radio luminosity exceeded (by 0.7 dex) that
expected for a SFR of $0.01 M_{\odot}$yr$^{-1}$. If their radio
luminosity was below that level, but above the radio-excess limit for
their estimated consensus SFR, they were deemed to be unclassifiable
in terms of radio excess (0.4 per cent of sources).

Finally, a small proportion of sources do not reach the radio-excess
selection threshold, but are clearly extended or multi-component radio
sources, inconsistent with simply being star-forming galaxies. Those
sources which are either multi-component sources associated through
the LOFAR Galaxy Zoo effort \citepalias{Kondapally2021} with a
physical size in excess of 80\,kpc, or single component sources with a
major axis size in excess of 80\,kpc and which also exceed the
resolved source threshold defined in \citet{Shimwell2019} by at least
a factor of 1.5, were deemed to be clearly extended. These sources
were added to the radio-excess sample if they were not already
included (just under 0.5 per cent of the sample).

The lower panels of Fig.~\ref{fig:SFRradio} show the ratio of measured
radio luminosity over that expected from the consensus SFR as a
function of redshift (left) and stellar mass (right); the horizontal
dashed lines show the expected relation for star-forming galaxies and
the radio-excess threshold, and the blue circles indicate again the
peak of the distribution at each redshift. It can be seen that there
is a weak variation of the population distribution with redshift, but
no consistent trend, and the distribution peak never moves more than
0.2 dex ($<1\sigma$) from the ridgeline value. Radio excess sources
are found across all redshifts. The apparent gradual decline in the
ratio with increasing redshift at $z>2.5$ may be due to an increasing
incompleteness in the classification of radiative-AGN at these
redshifts (see Sec.~\ref{sec:optAGN}), leading to an over-estimate in
the SFR of some sources.

Regarding stellar mass, it is immediately clear from the lower-right
panel of Fig.~\ref{fig:SFRradio} that the proportion of radio-excess
sources increases very strongly with mass, in particular for those
objects not selected to be radiative-mode AGN. This is the well-known
trend that, in the local Universe, the radio-loud AGN fraction shows a
very strong mass dependence \citep[e.g.][]{Best2005,Sabater2019}.
\citet{Kondapally2022} use this LoTSS-Deep sample to investigate the
cosmic evolution of this trend. Fig.~\ref{fig:SFRradio} also shows a
weak variation of the peak of the distribution of
observed-to-predicted radio luminosity with mass, with a consistent
trend of higher mass galaxies having on average a slightly higher
radio luminosity for a given SFR. This has been previously seen in the
radio luminosity to SFR relation \citep[e.g.][]{Gurkan2018,Smith2021},
but it remains unclear to what extent this is due to an intrinsic
mass-dependence of the amount of radio emission arising from star
formation, as opposed to the effect of a contribution from a
population of radio-weak AGN, more prevalent at higher stellar masses,
that fall below the selection limit for radio-excess sources.

Regardless, the variations in Fig~\ref{fig:SFRradio} are sufficiently
small (in both redshift and stellar mass) that the use of a single
SFR-radio relation does not significantly affect the selection of
radio-excess sources.

\section{Final radio source classifications, and dependencies}
\label{sec:classes}

In the previous sections, LoTSS-Deep sources have been identified as
either radiative-mode AGN or not, and either radio-excess sources or
not, with a small number of sources being unclassifiable in each
case. Here, these are combined to derive a final set of source
classifications.

\begin{itemize}
  \item Sources which are neither radiative-mode AGN nor radio-excess
    sources are classified simply as star-forming galaxies
    (SFGs). Note that this may include some quiescent galaxies (with
    SFRs below the stellar mass {\it vs} SFR main sequence) whose low
    redshift nevertheless allows the star formation to be detected by
    LOFAR.
  \item Sources which are radiative-mode AGN but which do not display
    a radio excess are radio-quiet AGN (RQAGN; including the
    radio-quiet quasars)
  \item Sources which are not radiative-mode AGN but do display a
    radio excess are the population of jet-mode AGN. Traditionally
    these sources are referred to as low-excitation radio galaxies
    (LERGs)
  \item Sources which are both radiative-mode AGN and radio-excess
    sources are sources such as radio-loud quasars (Type I or Type
    II). These are traditionally referred to as high-excitation radio
    galaxies (HERGs).
  \item Finally, any source which could not be reliably classified in
    either of the criteria was left as unclassified.
\end{itemize}

\begin{table*}
  \begin{center}
    \caption{\label{tab:nsource} The number of sources of each class
      in the LoTSS-Deep DR1 dataset.}
  \begin{tabular}{cccccc}
    \hline
Source classification   & ELAIS-N1   &Lockman Hole& ~~~Bo\"otes~~~ & ~~~Total~~~ &
Percentage \\
\hline
Star-forming galaxies          & 22720 & 21044 & 11916    & 55680 & 67.9 \\
Radio-quiet AGN                &  2779 &  2633 &  2030    &  7442 & 9.1 \\
Low-excitation radio galaxies  &  4287 &  5304 &  3158    & 12749 & 15.6 \\
High-excitation radio galaxies &   510 &   710 &   524    &  1744 & 2.1 \\
Unclassified                   &  1314 &  1471 &  1551    &  4336 & 5.3 \\
\hline
Total                          & 31610 & 31162 & 19179    & 81951 & 100\\
\hline
  \end{tabular}
  \end{center}
\end{table*}

Table~\ref{tab:nsource} shows the number of sources of each class in
each field. As can be seen, the majority population in LoTSS-Deep DR1
is the star-forming galaxies: these comprise just over two-thirds of
the total population, rising to over 70 per cent in the deepest field,
ELAIS-N1. Radio-quiet AGN contribute nearly 10 per cent of the total,
with the two radio-loud classes contributing around 18 per cent
between them, mostly as LERGs. Five per cent of the sources are
unclassified. Of these, around 3 per cent are the sources without host
galaxy identifications or redshifts for which no SED fitting could be
carried out, and the remaining 2 per cent are mostly fainter galaxies
for which the SED fitting algorithms either did not provide acceptable
fits or provided highly inconsistent results.

Table~\ref{tab:outputvalues} provides the first five lines of the
classification data for each source in ELAIS-N1, along with the
consensus mass and SFR measurements; the full catalogues for each
field are provided electronically. More extensive catalogues,
including the key outputs of each SED fitting code that were used to
derive these, are made available on the LOFAR Surveys website
(\url{lofar-surveys.org}).

\begin{table*}
  \begin{center}
    \caption{\label{tab:outputvalues} Classification results and
      consensus measurements for each source. The table shows the
      first five sources in ELAIS-N1: full catalogues are available
      electronically. Columns give the full source identifier, the
      radio ID number, the total 150\,MHz flux density (in Jy), the
      redshift, the final radiative-mode AGN classification (1$=$AGN,
      0$=$non-AGN, $-1$$=$unclassifiable), the logarithm of the
      consensus stellar mass (in solar masses), the logarithm of the
      consensus SFR (in solar masses per year), the radio excess (in
      dex), a flag to indicate extended radio sources (as defined in
      Sec.~\ref{sec:radAGN}; 1$=$extended, 0$=$compact), the final
      radio-AGN classfication (1$=$radio-AGN, 0$=$no radio excess,
      $-$1$=$unclassifiable), and the overall classification
      (SFR$=$star-forming galaxies; RQAGN$=$radio-quiet AGN;
      LERG$=$low-excitation (jet-mode) radio galaxy;
      HERG$=$high-excitation (quasar-mode) radio galaxy;
      Unc$=$unclassified. Values of $-$99 indicate where no
      measurement is available. }
  \begin{tabular}{ccccccccccc}
    \hline
    Source Name & Radio ID & $S_{\rm 150MHz}$ & $z$ & AGN & log$_{10}$(Mass) & log$_{10}$(SFR) & Radio excess & Extended & Radio & Overall \\
        &  & [Jy] & & class & [M$_{\odot}$] & [M$_{\odot}$/yr] & [dex] & & class & class \\
\hline
ILTJ155957.58+550052.4 & 0 & 0.000396 & 2.0437 & 0 & 11.62  & 2.22  &  0.31  &  0  &  0 & SFG\\
ILTJ155958.25+550105.3 & 1 & 0.000736 & 0.6697 & 0 & 11.00  & 1.58  &  0.15  &  0  &  0 & SFG \\
ILTJ155958.68+550534.6 & 2 & 0.000197 & 1.4289 & 0 & 11.58  & 1.16  &  0.79  &  0  &  1 & LERG \\
ILTJ155959.52+545751.0 & 3 & 0.000158 & 1.7777 & 0 & 11.20  & 1.71  &  0.32  &  0  &  0 & SFG \\
ILTJ160000.65+550723.3 & 4 & 0.000196 & 3.6960 & 1 & 11.42  & 2.87  & -0.13  &  0  &  0 & RQAGN \\
\dots & \dots & \dots & \dots & \dots & \dots & \dots & \dots & \dots & \dots \\ 
\hline
  \end{tabular}
  \end{center}
\end{table*}

\begin{figure*}
    \centering
    \includegraphics[width=\textwidth]{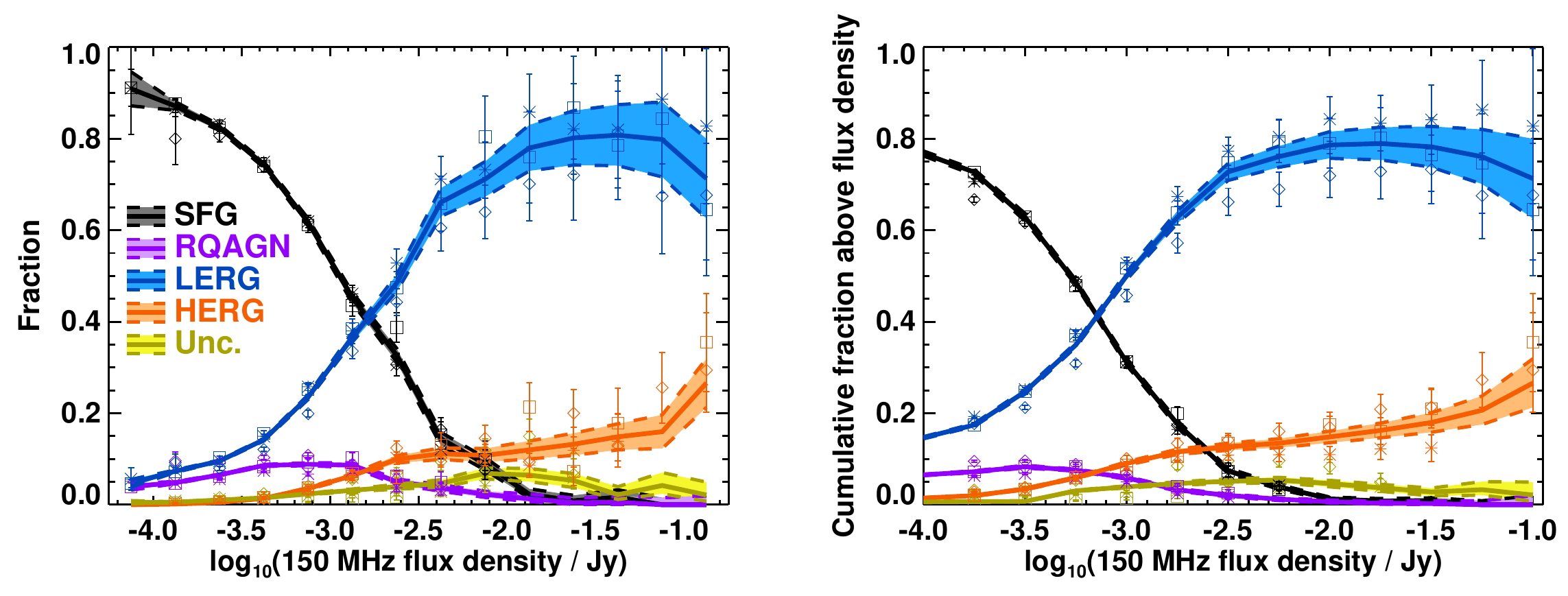}\vspace*{3mm}\\
    \includegraphics[width=\textwidth]{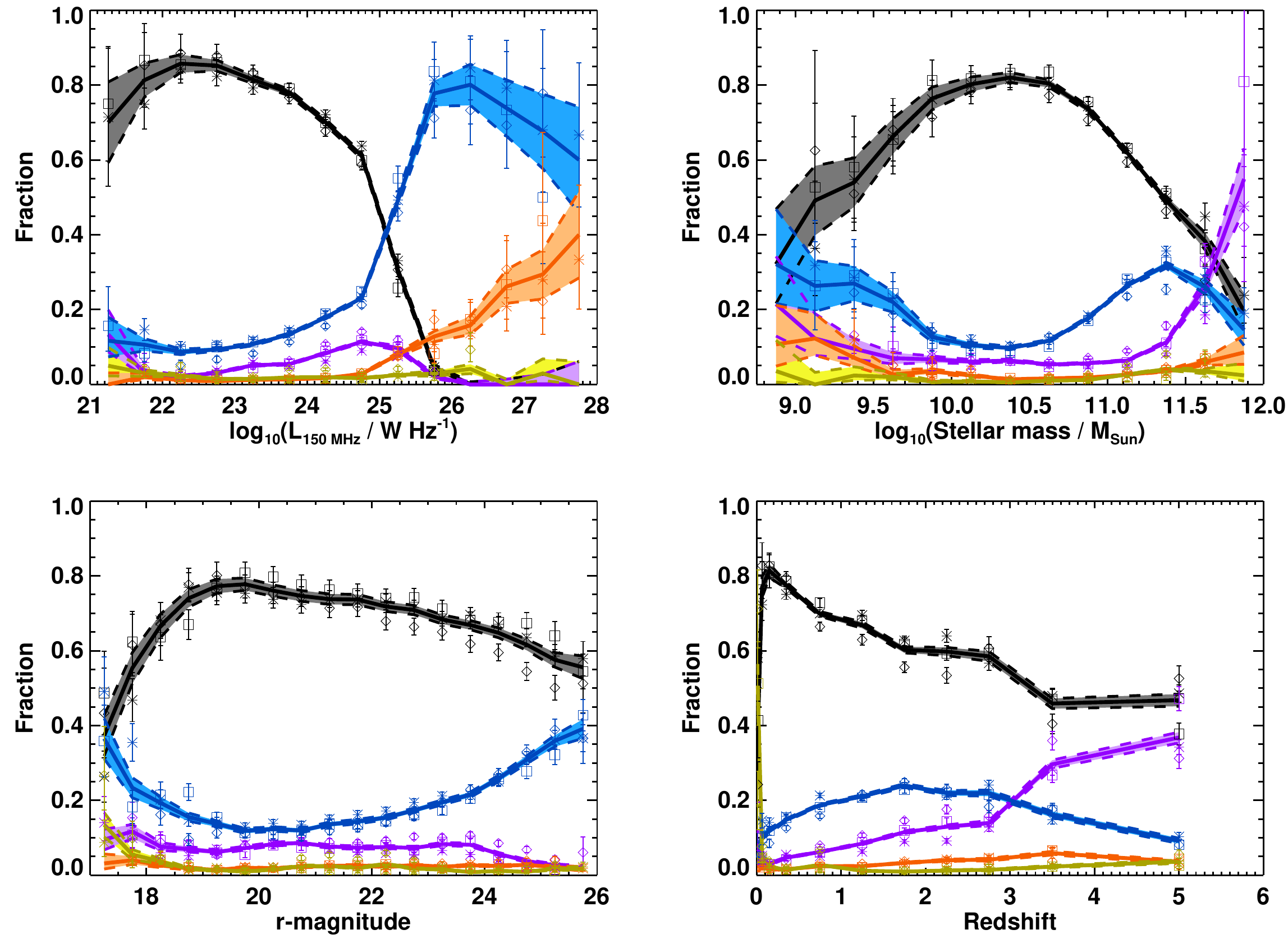}
    \caption{The fraction of sources of each different class
      (star-forming galaxies in grey; radio-quiet AGN in
        purple; low-excitation radio galaxies in blue; high excitation
        radio galaxies in orange; unclassifiable sources in yellow) as
        a function of radio flux density (upper panels; left gives
        fraction at a given flux density, and right gives cumulative
        fraction above a flux density), radio luminosity (middle
        left), stellar mass (middle right; for sources with $z<1.8$
        only -- see text), optical r-band magnitude (lower left) and
        redshift (lower right; out to a final bin of
          $4<z<6$). On each plot, the solid line for each class
        represents the derived fraction, and the shaded region
        indicates the calculated uncertainty. The open symbols show
        the values derived from each individual field (square $=$
        ELAIS-N1; asterisk $=$ Lockman Hole; diamond $=$ Bo\"otes),
        where there are at least 5 sources from that field in the
        given bin, and demonstrate the broad agreement between
        fields. Note that the rise of the radio-quiet AGN population
        at the highest stellar masses is probably an artefact of
        larger mass uncertainties for these sources; see text for
        details.}
      \label{fig:sourcepops}
\end{figure*}

Figure~\ref{fig:sourcepops} shows the distribution of the different
classes of source as a function of various properties of the host
galaxy. The top panels show the distribution with respect to the
150-MHz flux density: the left panel shows the fraction at a given
flux density, and the right panel shows the cumulative fraction above
a given flux density. The population is dominated by radio-loud AGN
above flux densities of about a mJy. The bulk of these are the LERGs,
but with the fraction of HERGs beginning to rise at the highest flux
densities, where the coverage of the sample begins to run out due to
lack of sky area for these rarer bright sources. This rise of the HERG
population is seen even more starkly in the middle left panel, which
shows the distribution as a function of radio luminosity, and is in
line with expectations from the relative luminosity functions of these
two populations \citep[e.g.][]{Best2012,Best2014}. At lower flux
densities (and below 150\,MHz luminosities of around
$10^{25}$W\,Hz$^{-1}$), star-forming galaxies take over the sample and
quickly become the dominant population, accounting for over 90 per
cent of sources at the limiting flux density reached in ELAIS-N1 (and
more than 75 per cent of the cumulative population above $S_{\rm
  150MHz} \approx 100\mu$Jy). The switch between a star-formation
dominated population and a radio-loud AGN dominated population occurs
at around $S_{\rm 150 MHz} \approx 1.5$\,mJy, which is fully
consistent with the switch point at higher frequency of $S_{\rm 1.4
  GHz} \approx 200 \mu$Jy \citep[found by][]{Smolcic2017b} or $S_{\rm
  1.4 GHz} \approx 250 \mu$Jy \citep[found by][]{Padovani2016},
considering the typical radio spectral index of these sources.

At all flux densities below a few mJy there is a significant
population of radio-quiet AGN, accounting for just under 10 per cent
of all sources over the 100\,$\mu$Jy to 1\,mJy flux density
range. This is slightly lower than the fraction found in observations
at higher frequencies: early work by \citet{Simpson2006} suggested
that 20 per cent of sources with 100\,$\mu$Jy $\lta S_{\rm 1.4GHz}
\lta 300$\,$\mu$Jy are radio-quiet AGN, while the COSMOS 3GHz work of
\citet{Smolcic2017b} indicated between 15 and 20 per cent (as
determined from the 70 per cent subset of their `High Luminosity AGN'
sample that shows no radio excess). The origin of this difference is
not completely clear. It may be related to different implementations
of the radio-loud to radio-quiet separation, but more likely is
associated with the radio-quiet AGN having a flatter spectral index
than star-forming galaxies (e.g.\ due to a greater proportional
contribution of flatter-spectrum core emission) and therefore lesser
prominence at the lower frequencies probed by LOFAR. Given the
steepness of the radio source counts, a difference of only
$\approx$0.2 in spectral index between star-forming galaxies and
radio-quiet AGN would decrease the proportion of radio-quiet AGN in
the sample by about a factor of 2; LOFAR studies of radio-quiet
quasars provide evidence in support of such flatter spectral indices
\citep[e.g.][]{Gloudemans2021}.

The additional panels of Fig.~\ref{fig:sourcepops} show the
distribution of source classes as a function of redshift, stellar mass
and optical magnitude. Note the strong rise of unclassified sources at
$z<0.1$; low SFRs for these galaxies can also lead to ambiguous radio
excesses, while in addition the aperture photometry and aperture
corrections used for the LoTSS Deep Field photometry
\citepalias{Kondapally2021} are not optimised for these low redshifts,
and resulting errors will affect the SED fitting. At these redshifts,
it is in any case better to use the shallower, wider-area LoTSS
surveys. All populations are seen over the full range of optical
magnitudes. As expected, the LERG population shows increasing
importance at higher stellar masses (note that this panel only
includes redshifts $z<1.8$ as mass estimates become increasingly less
reliable at higher redshifts). The radio-quiet AGN show a dramatically
increasing importance at stellar masses above $10^{11.5} M_{\odot}$,
but this is likely to be an artefact, driven by larger mass
uncertainties for these sources due to the potential AGN contributions
to their spectra: the number of sources at these very highest masses
is relatively low, and so a few sources scattered up to high masses
due to wider uncertainties on their masses, or due to errors in the
photometric redshifts pushing them to higher redshift (and hence
higher luminosity and mass), can artificially dominate the
population. Interestingly, star-forming galaxies are seen across the
full range of redshifts studied; this indicates that the LoTSS-Deep
sample is not only able to study normal star-forming galaxies in the
low and moderate redshift Universe, but also to select starbursting
galaxies in the early Universe.

All of these results are broadly consistent across the three fields
(indicated by the open symbols in Fig.~\ref{fig:sourcepops}). In
Sec.~\ref{sec:optAGN}, the threshold levels for selection of
radiative-mode AGN were set slightly differently in Bo\"otes than the
other two fields, based on the typically higher $f_{\rm AGN}$ values
found for the known spectroscopic and X-ray AGN and colour-selected
probable AGN. The consistency of the classifications between fields in
Fig.~\ref{fig:sourcepops} gives confidence that this variation in
thresholds is indeed appropriate. The remaining variations are
consistent with what might be expected from cosmic variance, and
indicate the importance of combining the multiple fields in order to
overcome these effects, as well as to build a large statistical sample
of sources.

\section{Comparisons with simulated sky models}
\label{sec:modelcomp}

Radio sky simulations provide a valuable tool for predicting the
populations of radio sources that will be observed in a given
survey. In addition to the planning of future radio surveys
\citep[e.g.][]{Norris2013} or predictions of parameter constraints
achievable with those \citep[e.g.][]{Raccanelli2012,Harrison2016},
these simulations are a valuable tool in assessing the completeness of
different radio surveys \cite[e.g.][]{Hale2022}, or in generating
random samples for clustering analyses \citep[e.g.][]{Siewart2020}.
The two most widely used radio sky simulations in the literature are
the SKA Design Study (SKADS) Simulated Skies \citep{Wilman2008} and
the more recent Tiered Radio Extragalactic Continuum Simulation
\citep[T-RECS;][]{Bonaldi2019}. 

The starting point for these simulations is the measured luminosity
functions of different source populations, and their cosmic evolution,
which has typically been measured out to intermediate redshifts. The
luminosity functions are then extrapolated to lower luminosities
(lower flux densities), evolved out to higher redshifts, and
potentially converted to a different observed frequency. Comparison of
the predictions of these models against new deep observations such as
the LoTSS Deep Fields provides a critical test of the assumptions that
go into the radio sky simulations, and an opportunity to revise and
improve these.

SKADS provides simulated predictions for four different radio source
populations: star-forming galaxies, radio-quiet AGN, and two
populations of radio-loud AGN. The two radio-loud AGN populations
represent a low-luminosity and a high-luminosity component that
\citet{Wilman2008} associated with the FRI and FRII morphological
sub-populations \citep{Fanaroff1974}, but which also map reasonably
well onto the LERG and HERG classifications, respectively, used in
this paper. Thus, all four radio source populations can be directly
compared between the SKADS simulations and the LoTSS-Deep data. The
radio-loud AGN population in T-RECS is constructed from luminosity
functions for steep- and flat-spectrum radio sources together with BL
Lac objects: these do not map onto the radio-AGN subclasses considered
here, so comparisons with T-RECS can only be made with the radio-loud
AGN population as a whole. T-RECS also includes predictions for SFGs,
but does not include a separate radio-quiet AGN population: instead,
T-RECS assumes that the radio emission of radio-quiet AGN is dominated
by the on-going star-formation and thus that the radio-quiet AGN are
encompassed within the star-forming population.

\begin{figure*}
    \centering
    \includegraphics[width=0.95\textwidth]{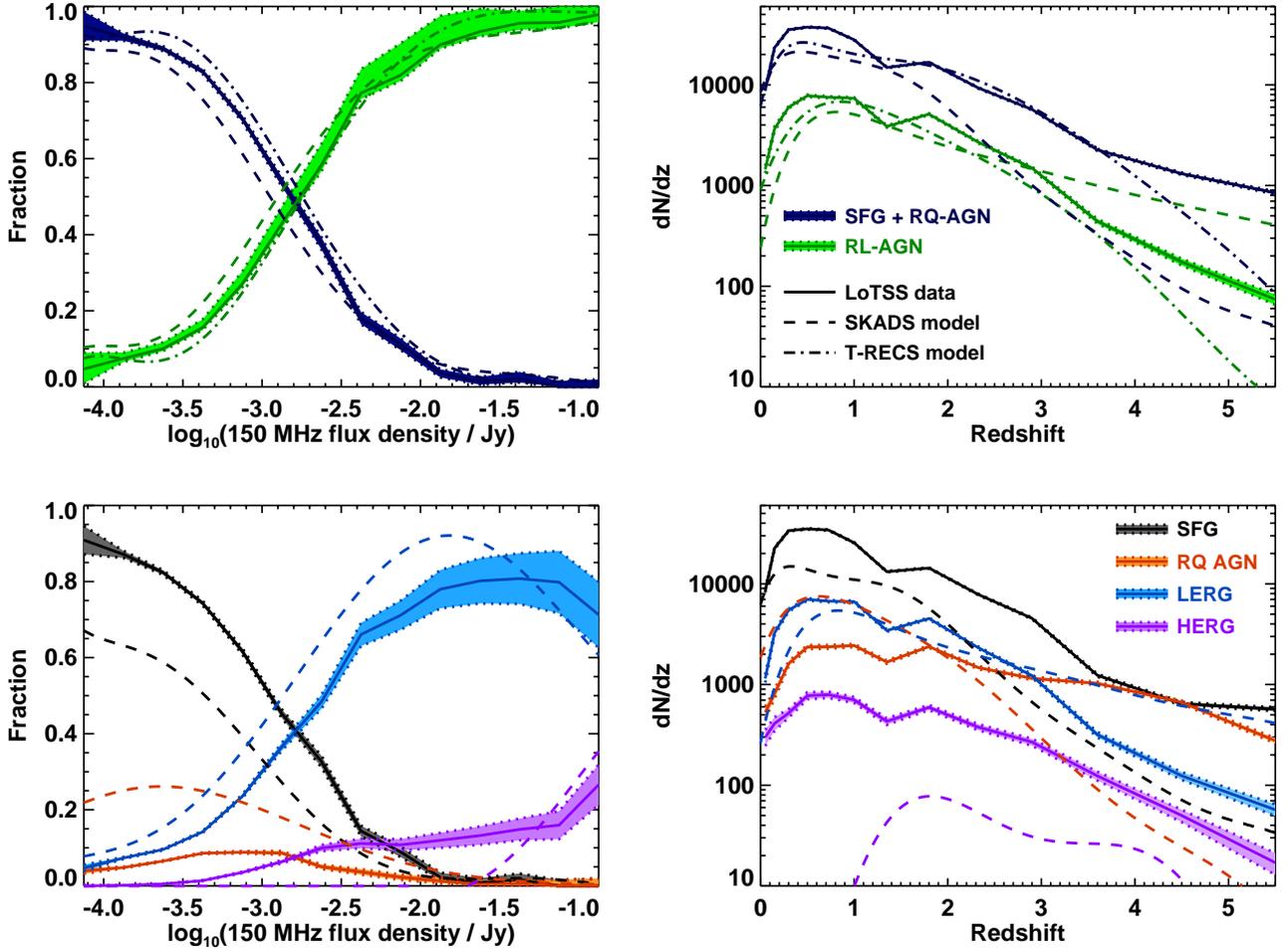}
    \caption{A comparison of the radio source population fractions as
      a function of 150\,MHz flux density (left panels) and the
      redshift distribution of radio sources (right panels) between
      the LoTSS-Deep data (solid lines and shaded regions) and the
      simulated sky predictions from SKADS \citep[][dashed
        lines]{Wilman2008} and T-RECS \citep[][dot-dash
        lines]{Bonaldi2019}. The upper panels show the populations
      split just into star-forming galaxies plus radio quiet AGN
      (blue) versus radio-loud AGN (green), which can be compared
      against both SKADS and T-RECS simulations. The lower panels
      compare the four sub-populations against the SKADS simulation
      predictions; note that the separation of the two SKADS
      radio-loud classes does not map precisely onto the HERG/LERG
      classification used in this paper, although it is reasonably
      similar (see text).}
    \label{fig:modelcomp}
\end{figure*}

For both the SKADS and T-RECS simulations, a predicted source
population was extracted over a randomly-located sky area
corresponding to each of the three LoTSS Deep Fields. The
  simulations include sources to well below the flux limits of the
  observation and so, to replicate the observations, the LoTSS-Deep
completeness simulations of \citet{Kondapally2022} and
\citet{Cochrane2023} were used to determine the probability that each
simulated source would be detected, and the source was randomly
included in, or excluded from, the simulated catalogue in accordance
with that probability.  Figure~\ref{fig:modelcomp} shows how the
resultant simulated samples compare against the LoTSS-Deep data in
both flux density (left panels) and redshift (right panels). Note that
the small dip in the redshift distribution of all LoTSS-Deep
populations over $1.0 < z < 1.5$ is due to an aliasing effect in the
photometric redshifts, particularly in the ELAIS-N1 and Lockman Hole
fields, probably due to the lack of H-band data; this is discussed in
more depth in \citet{Cochrane2023}, but is not a significant issue for
the analysis in the current paper.

The upper panels of Figure~\ref{fig:modelcomp} show the simulation
{\it vs} data comparison for a simple split into the two T-RECS source
populations: star-forming galaxies plus radio-quiet AGN, against
radio-loud AGN (HERGs + LERGs). Note that as well as allowing a
comparison against both T-RECS and SKADS, this population split is
arguably the most robustly determined in the LoTSS-Deep dataset, as it
depends only on the presence or absence of a radio-excess rather than
the (more difficult to establish) evidence for a radiative AGN. These
upper panels show that both T-RECS and SKADS describe fairly well the
transition between these two populations with decreasing radio flux
density. T-RECS also provides an accurate match to the redshift
distribution out to redshift $z \sim 4$, beyond which the simulated
source counts fall below those measured in the data; it is not clear
whether this is a shortcoming of the simulation, or whether the
photometric redshifts of the highest redshift sources become less
reliable. The SKADS simulations also match the data reasonably well
out to redshift $z \sim 2$, but thereafter they over-predict the
number of radio-loud AGN and under-predict the star-forming galaxy
population.

The lower panels of Figure~\ref{fig:modelcomp} provide further
analysis of the SKADS simulations, split into the four
sub-populations. Here, signficant differences are observed between the
simulated and observed datasets. First, SKADS underpredicts the number
of SFGs by a factor $\approx 2$ at all redshifts $z \gta 0.2$. This is
a result which has previous been established
\citep[e.g][]{Bonaldi2016,Smolcic2017}; \citet{Hale2022} use a
`modified SKADS' model where they double the number of star-forming
galaxies. Second, SKADS substantially over-predicts the number of
radio-quiet AGN at lower redshifts and lower flux densities compared
to the observations. Although it cannot be excluded that this is due
to misclassification of faint radio-quiet AGN as star-forming galaxies
in the observational data, a more likely explanation is that, as
discussed earlier, this is due to an assumed radio spectral index of
0.7 for the radio-quiet AGN; a flatter spectral index (or
  curved spectral shape due to low-frequency absorption) would lead
to a lower prevalence of these sources at the low frequencies of the
LoTSS-Deep data. The combination of fewer SFGs and more RQAGN gives
rise to the good agreement at low redshifts in the upper panel. For
the radio-loud AGN, the difference in the high redshift number counts
comes primarily from an over-prediction of the LERG population; the
high redshift evolution of these sources was unknown at the time of
the SKADS simulations, and so was assumed to be flat beyond $z \sim
0.7$; recent works \citep[e.g][]{Kondapally2022} show this to be a
reasonable assumption out to $z \sim 2$, but with indications of a
decline between $2.0<z<2.5$, suggesting a breakdown of the SKADS
assumptions.

In conclusion, while the SKADS simulations have been very successful
in producing simulated radio skies, datasets such as LoTSS-Deep which
probe new parameter space are revealing the shortcomings in
our understanding 15 years ago when those simulations were first
produced. The more modern T-RECS simulations provide a better match to
the current dataset, but would be enhanced by the explicit inclusion
of a radio-quiet AGN dataset, since the assumption that the radio
emission of these sources is entirely produced through star-formation
is known not to be true \citep[see e.g.][]{Macfarlane2021}.
Furthermore, explicit separation of the radio-loud population into
HERG and LERG components in T-RECS would be a valuable addition and
allow more detailed comparison of the simulation performance.

\section{Summary}
\label{sec:concs}

The LoTSS Deep Fields are the widest deep radio survey ever
undertaken. The LoTSS-Deep first data release, comprising
$\approx$80,000 radio sources, is already an order of magnitude larger
than previous radio source samples at this depth. The final LoTSS-Deep
sample will detect $> 250,000$ radio-selected sources over a
35\,deg$^2$ region of sky, split into four different fields to largely
overcome cosmic variance. Extensive multi-wavelength photometry from
the UV to the far-IR in each field facilitates a huge range of
scientific exploitation.

In this paper, a combination of four different SED fitting codes has
been applied to the multi-wavelength photometry of each of the
LoTSS-Deep DR1 sources. Two of the four codes (\cigale\ and
\agnfitter) include an AGN component in their SED modelling, and these
offer an estimate of the AGN contribution to the overall galaxy
SED. The other two codes (\magphys\ and \bagpipes) do not include AGN
components, but offer more comprehensive coverage of the parameter
space of the stellar component, and therefore are able to provide more
accurate results for galaxies without AGN contributions. By combining
the AGN fractional contributions estimated by \cigale\ and
\agnfitter\ with the relative fitting ability of these two codes
compared against \magphys\ and \bagpipes, those galaxies with an AGN
contribution to their SED are identified.

Consensus stellar masses and star-formation rates are determined for
each galaxy. For the galaxies without AGN contributions, these are
generally based on the \magphys\ and \bagpipes\ results, which show
excellent overall agreement with each other. For those which do show
an AGN contribution to their spectra, the \cigale\ results are
primarily adopted, as \cigale\ is shown to provide more reliable
estimates than \agnfitter.

The consensus star-formation rates are used to determine a
relationship between 150\,MHz radio luminosity and star-formation
rate, using a `ridgeline' approach to minimise bias from both radio
selection effects and weak radio-AGN contributions. The determined
relation is $\log_{10} L_{\rm 150 MHz} = 22.24 + 1.08 \log_{10}({\rm SFR})$,
where $L_{\rm 150 MHz}$ is in units of W\,Hz$^{-1}$ and SFR in units
of $M_{\odot}$\,yr$^{-1}$. This is in very good agreement with
previous literature studies.  Radio-excess sources are then identified
as those sources which show at least 0.7 dex (corresponding to
$\approx 3\sigma$) more radio emission than would be expected based on
the star formation rate.

Using these results, the LoTSS Deep Field sources are then classified
into four classes: (i) star-forming galaxies, which show neither any
evidence for an AGN in their optical/IR SED nor a radio-excess; (ii)
radio-quiet AGN, which do have an AGN contribution to their optical/IR
SED, but show no radio excess; (iii) low-excitation radio galaxies
(jet-mode radio-AGN), which show a radio excess but no optical/IR AGN
signatures; (iv) high-excitation radio galaxies which show both AGN
emission in their optical/IR SED and a radio excess. Less than 5 per
cent of the sources are unable to be classified. Overall, over
two-thirds of the sources in the LoTSS Deep Fields are star-forming
galaxies, around 16 per cent are LERGs, just under 10 per cent are
radio-quiet AGN, and 2 per cent are HERGs.  The three LoTSS Deep
Fields show strong agreement in their source populations, despite
significant differences in the input multi-wavelength photometric
data.

The star-forming galaxies dominate the population below flux densities
of $S_{\rm 150 MHz} \approx 1$\,mJy, accounting for $\approx$90 per
cent of the sources close to the flux limit of the deepest field,
$S_{\rm 150 MHz} \lta 100 \mu$Jy. In terms of luminosity, the
star-forming galaxies become the largest population below $L_{\rm 150
  MHz} \approx 10^{25}$W\,Hz$^{-1}$. At higher flux densities, and
higher luminosities, the LERGs are the dominant population. The
proportion of HERGs begins to rise significantly at the very highest
flux densities and luminosities, but the LoTSS Deep Fields do not
cover enough sky area to probe the regime where these become the
dominant population.

Star-forming galaxies are observed across all redshifts, ranging from
normal star-forming galaxies in the nearby Universe to extreme
starbursting systems at $z>4$. They are also observed across a wide
range of optical magnitudes and stellar masses, peaking at around
$10^{10.5}$ solar masses, typical of galaxies towards the upper end of
the star-forming main sequence. The proportion of radio-quiet AGN
rises noticeably towards higher redshifts; it also rises sharply
towards the highest stellar masses, but this is likely to be an
artefact of the steep stellar mass function coupled with larger
uncertainties on the stellar masses of this population. The LERG
population reaches its peak importance at redshifts 1 to 3; however,
the proportion of LERGs is smaller than that of the star-forming
galaxies at all redshifts, stellar masses and optical magnitudes.

The observed populations are compared against the prediction of the
SKADS and T-RECS radio sky simulations. SKADS is shown to underpredict
the star-forming galaxy population by a factor $\approx 2$ across all
redshifts. It over-predicts the proportion of radio-quiet AGN in the
sample. This is likely to be due to the assumption of a
  radio spectral index of $\alpha = 0.7$ for these sources: a flatter
spectral index, as indicated by recent LOFAR observations of
radio-quiet quasars, would reduce the prevalence of these sources in
these low-frequency observations. Finally, SKADS over-predicts the
numbers of LERGs at redshifts $z > 2$, as it does not account for the
negative cosmic evolution of this population at high redshift
beginning to be observed in the latest datasets. T-RECS provides a
good match to the star-forming and radio-loud AGN populations, but its
lack of a separate radio-quiet AGN population is a significant
shortcoming.

The classifications, stellar masses and SFRs derived in this paper
form a vital input to many other studies using the LoTSS Deep Fields
first data release \citep[][and
  others]{Smith2021,Bonato2021,Kondapally2022,McCheyne2022,Mingo2022,Cochrane2023},
and the techniques developed to derive these can be applied to future
data releases of the LoTSS Deep Fields.  Many advances continue to be
made in the LoTSS Deep Fields that, in addition to new deeper radio
data, will improve classifications still further. Over the next 5
years, the WEAVE-LOFAR survey \citep{Smith2016} will obtain around a
million optical spectra of LOFAR sources, including all sources
detected in the LoTSS Deep Fields, using the new William Herschel
Telescope (WHT) Enhanced Area Velocity Explorer (WEAVE) multi-object
spectrograph \citep{Jin2022}. WEAVE-LOFAR will provide spectroscopic
redshifts for the vast majority of the star-forming galaxies,
radio-quiet AGN and HERGs (especially at lower redshifts) due to their
strong emission lines, removing one of the largest uncertainties in
the SED fitting. It may be possible to obtain spectroscopic redshifts
for LERGs from weaker lines or continuum features, and even where this
is not the case, the confirmed absence of strong emission lines and
AGN features will add confidence to the reliability of the photometric
redshifts. For many sources, WEAVE-LOFAR will also improve source
classifications through either emission line diagnostics, or emission
line to radio flux ratios \citep[cf.][at lower
  redshifts]{Best2012}. Future imaging of these fields at 0.3-arcsec
resolution, by including the international LOFAR baselines
\citep[cf.][]{Morabito2022,Sweijen2022}, will further improve source
classification by allowing compact radio cores (AGN), kpc-scale
star-forming regions, and small-scale core-jet radio sources to be
distinguished by their radio morphology in these fields
\citep{Morabito2022}. A comparison between the SED-determined
classifications and those from high resolution radio morphology will
be very interesting.

The final LoTSS-Deep sample, imaged with sub-arcsec radio resolution
and coupled with high-resolution optical spectroscopy for each source,
will represent an extremely powerful resource for studies of the
evolution of galaxies and AGN.

\section*{Acknowledgements}

PNB, JS and RK are grateful for support from the UK STFC via grant
ST/R000972/1 and ST/V000594/1. RK acknowledges support from an STFC
studentship via grant ST/R504737/1. MJH and DJBS acknowledge support
from STFC via grant ST/V000624/1. BM acknowledges support from STFC
under grants ST/R00109X/1, ST/R000794/1, and ST/T000295/1. WLW
acknowledges support from the CAS-NWO programme for radio astronomy
with project number 629.001.024, which is financed by the Netherlands
Organisation for Scientific Research (NWO). KJD acknowledges funding
from the European Union’s Horizon 2020 research and innovation
programme under the Marie Sk\l{}odowska-Curie grant agreement
No. 892117 (HIZRAD). CLH acknowledges support from the Leverhulme
Trust through an Early Career Research Fellowship.  KM is supported by
the Polish National Science Centre grant UMO-2018/30/E/ST9/00082. MB
and IP acknowledge support from INAF under the SKA/CTA PRIN `FORECaST'
and the PRIN MAIN STREAM `SAuROS' projects. MB also acknowledges
support from the Ministero degli Affari Esteri e della Cooperazione
Internazionale - Direzione Generale per la Promozione del Sistema
Paese Progetto di Grande Rilevanza ZA18GR02. MJJ acknowledges support
from the Oxford Hintze Centre for Astrophysical Surveys. LKM is
grateful for support from the UKRI Future Leaders Fellowship (grant
MR/T042842/1). RJvW acknowledges support from the VIDI research
programme with project number 639.042.729, which is financed by the
Netherlands Organisation for Scientific Research (NWO). 
  We thank the anonymous referee for helpful comments.

This paper is based on data obtained with the International LOFAR
Telescope (ILT) under project codes LC0\_019, LC2\_024, LC2\_038,
LC3\_008, LC4\_008, LC4\_034 and LC10\_012. LOFAR
\citep{vanHaarlem2013} is the Low Frequency Array designed and
constructed by ASTRON. It has observing, data processing, and data
storage facilities in several countries, that are owned by various
parties (each with their own funding sources), and that are
collectively operated by the ILT foundation under a joint scientific
policy. The ILT resources have benefitted from the following recent
major funding sources: CNRS-INSU, Observatoire de Paris and
Universit{\'e} d'Orl{\'e}ans, France; BMBF, MIWF-NRW, MPG, Germany;
Science Foundation Ireland (SFI), Department of Business, Enterprise
and Innovation (DBEI), Ireland; NWO, The Netherlands; The Science and
Technology Facilities Council, UK; Ministry of Science and Higher
Education, Poland. This research made use of the LOFAR-UK computing
facility located at the University of Hertfordshire and supported by
STFC [ST/P000096/1],

For the purpose of open access, the author has applied a Creative
Commons Attribution (CC BY) licence to any Author Accepted Manuscript
version of this paper.

\section*{Data Availability}

The data used in this paper come from the LoTSS Deep Fields Data
Release 1. The radio images and radio catalogues are presented by
\citet{Tasse2021} and \citet{Sabater2021}, and were made publicly
available through both the Centre de Donn\'ees astronomiques de
Strasbourg (CDS) and through the LOFAR Surveys website at
\url{https://lofar-surveys.org/deepfields.html}. The multi-wavelength
photometric catalogues and photometric redshifts come from
\citet{Kondapally2021} and \citet{Duncan2021} respectively, both of
which are also available through CDS and the LOFAR Surveys website.
For each field, a table of classifications, stellar masses and SFRs is
made available electronically as part of this paper. Furthermore, the
adapted input photometric catalogue developed in
Sec.~\ref{sec:sedcats} for the SED fitting, and a table of the key SED
fitting results from Secs.~\ref{sec:optAGN},~\ref{sec:consensus}
and~\ref{sec:radAGN} have been made available on
\url{https://lofar-surveys.org/deepfields.html}. More extensive SED
fitting results from each code can be made available upon reasonable
request to the corresponding author.




\bibliographystyle{mnras}
\bibliography{biblio} 


\appendix

\section{Uncertainties on stellar masses and SFRs}
\label{app:masserrors}

As discussed in the main text, no attempt is made to derive stellar
mass or SFR uncertainties on a source-by-source basis: any reader
interested in individual sources can examine the results of all of the
different SED codes, provided in the extended tables on the
lofar-surveys.org website, and make their own assessment of the
relevant systematic and statistical errors. Instead, this appendix
examines typical uncertainties that can broadly be considered.

\begin{figure}
    \centering
    \includegraphics[width=\columnwidth]{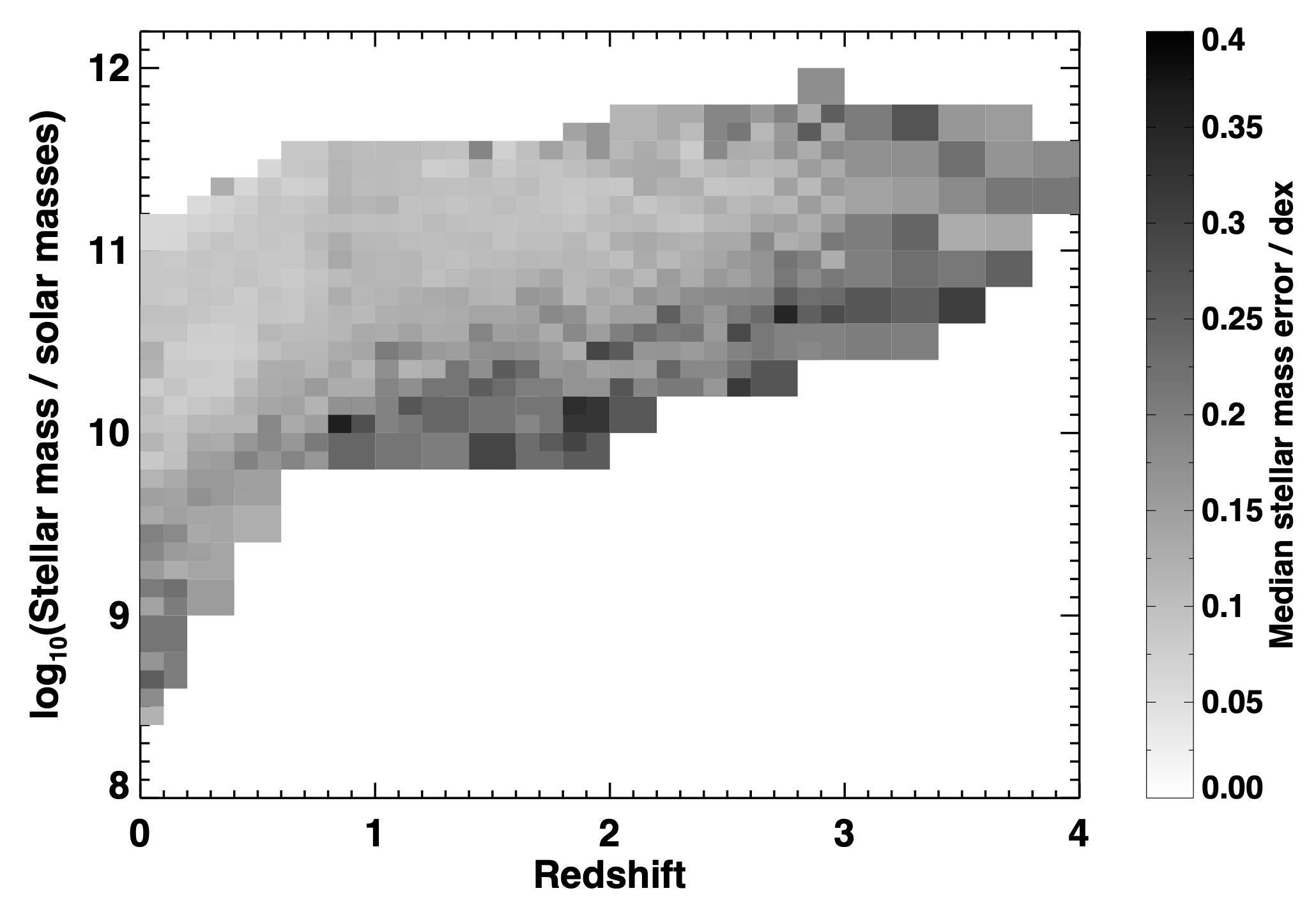}
    \caption{Typical uncertainties on estimates of stellar mass, as a
      function of both stellar mass and redshift, for non-AGN in the
      ELAIS-N1 field. At lower redshifts and higher stellar masses,
      the uncertainty is generally $\sim 0.1$dex, but this increases
      towards lower stellar masses and for redshifts $z>2$.}
    \label{fig:masserror}
\end{figure}

For sources not identified as radiative-mode AGN, the consensus
stellar masses are derived using the \magphys\ and
\bagpipes\ results. For each source, the difference between the
outputs of the two codes provides an indication of systematic
uncertainties, while the confidence intervals provided by each code
give an estimate of statistical uncertainty. As a broad guide to the
dominant uncertainty, for each source the higher of these two values
is considered. Figure~\ref{fig:masserror} then shows the median of
this value for all galaxies within a given bin in redshift--mass
space. As can be seen, the calculated median uncertainties are
typically $\lta 0.1$ dex at lower redshifts and higher masses. As
expected, they increase towards lower masses and towards higher
redshifts, in both cases due to the galaxies being fainter and
therefore having lower signal-to-noise photometric measurements in the
SED fitting.

The uncertainties in Fig.~\ref{fig:masserror} can broadly be
categorised in four different ranges of parameter space, with
empirical estimates of the uncertainty possible for each:

\begin{itemize}
\item Higher mass, lower redshift: specifically $\log_{10} M_* \ge (9.7 +
  2.5 \log_{10}(1+z))$ and $z \le 2$. Here the uncertainty on stellar mass
  is fairly constant at $\Delta M_* \approx 0.1$ dex.
\item Higher mass, higher redshift: specifically $\log_{10} M_* \ge (9.7 +
  2.5 \log_{10}(1+z))$ and $z > 2$. Here the uncertainty increases with
  increasing redshift and can be approximated as $\Delta M_* \approx
  0.05z$ dex.
\item Lower mass, lower redshift: specifically $\log_{10} M_* < (9.7 +
  2.5 \log_{10}(1+z))$ and $z \le 2$. Here the uncertainty increases with
  decreasing mass and increasing redshift, broadly as $\Delta M_*
  \approx 0.1 + 0.08 (1+z) (9.7 + 2.5\log_{10}(1+z) - \log_{10} M_*)$ dex.
\item Lower mass, higher redshift: specifically $\log_{10} M_* < (9.7 + 2.5
  \log_{10}(1+z))$ and $z > 2$. The relative high uncertainties here match
  on to the lower redshift and higher mass regimes: $\Delta M_*
  \approx 0.05z + 0.24 (9.7 + 2.5\log_{10}(1+z) - \log_{10} M_*)$ dex.
\end{itemize}

Uncertainties on the stellar masses of the radiative AGN are harder to
estimate in this manner, as mass estimates are derived from the two
\cigale\ fits, and these are likely to be subject to related
systematic errors. Comparing the confidence intervals of the
\cigale\ fits with those of the non-AGN in the same redshift-mass bin,
the statistical uncertainties of the radiative-mode AGN are on average
20 per cent larger than those of the non-AGN; this sets a lower limit
to the mass uncertainty estimate, although it is likely that the
systematic errors will also be larger in cases where the AGN
contributes significantly to the optical to near-IR spectrum.

\begin{figure}
    \centering
    \includegraphics[width=\columnwidth]{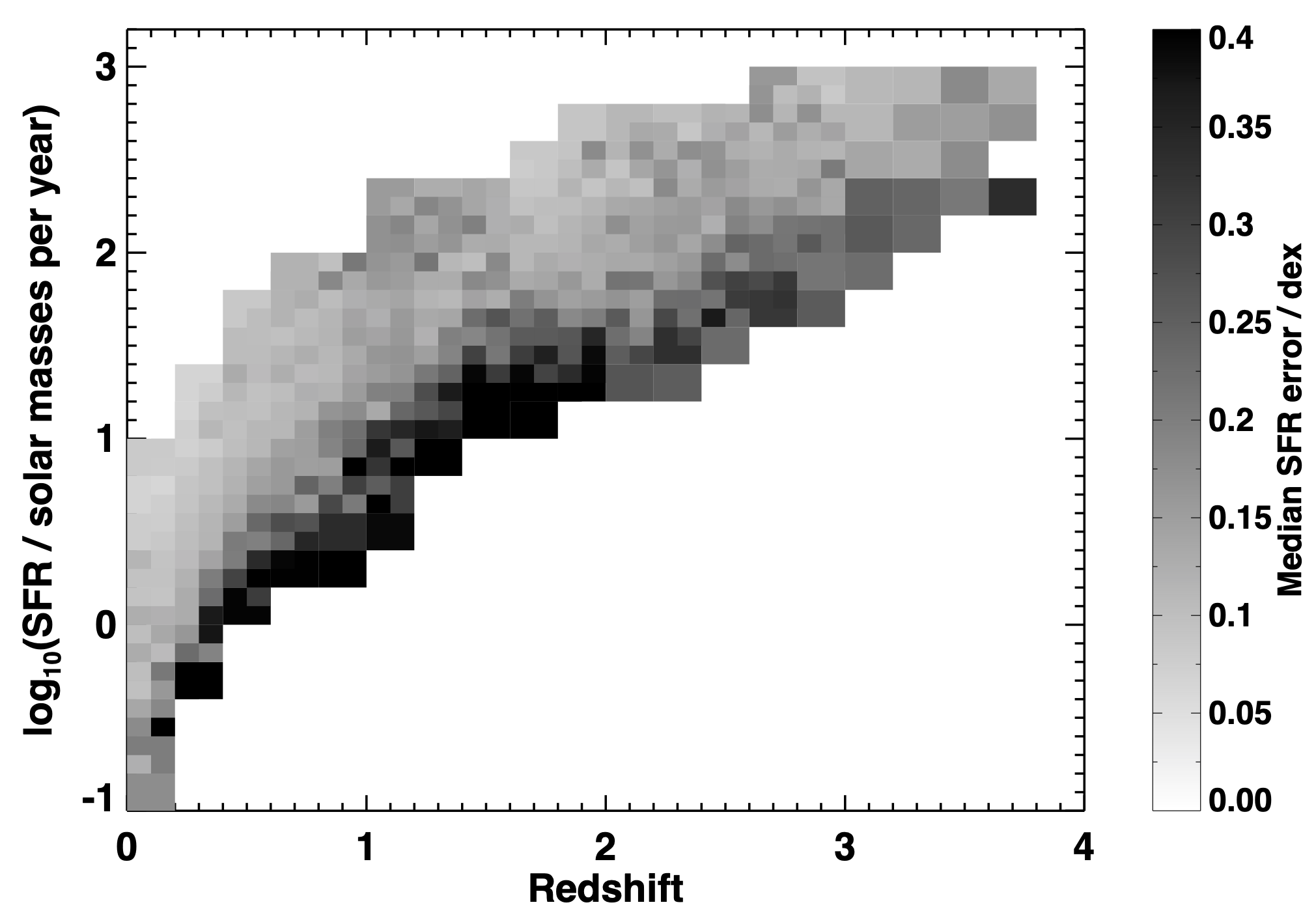}
    \caption{Typical uncertainties on estimates of SFR, as a function
      of both SFR and redshift, for non-AGN in the ELAIS-N1 field. SFR
      uncertainties increase dramatically for the few per cent of
      sources at the lowest SFRs at each redshift; above that they
      have little dependence on SFR, but increase gradually with
      increasing redshift.}
    \label{fig:sfrerror}
\end{figure}

A similar approach can be followed to estimate the typical
uncertainties on the consensus SFR estimates. Fig.~\ref{fig:sfrerror}
shows the result, split into bins of SFR and redshift. In this case it
is apparent that the SFR estimates are generally robust until the very
lowest SFRs at any redshift are reached (at most a few per cent of
objects), where the uncertainties increase dramatically. For the vast
majority of the population at higher SFRs, there is no strong
dependence of the SFR uncertainty (in dex) on the measured SFR, but a
clear trend for the uncertainty to increase with redshift, from
$\approx 0.1$ dex at $z \sim 0$ up to 0.15 dex by $z \sim 1$ and 0.2
dex by $z=3$. This can be empirically approximated as $\Delta$(SFR)
$\approx 0.1 \times (1+z)^{0.5}$ dex. The contributions
  to this uncertainty from differences between codes and from the
  statistical uncertainties within individual codes are comparable in
  size.

It should be emphasized again that these empirical relations are only
intended to provide a guide to the approximate stellar mass and SFR
uncertainties, and do not represent reliable values on a
source-by-source basis.

\bsp	
\label{lastpage}
\end{document}